\documentclass[12pt,a4paper]{article}

\usepackage[margin=0.8in]{geometry}

\usepackage[utf8]{inputenc}
\usepackage{amsmath}
\usepackage{etoolbox}
\usepackage{amsfonts}
\usepackage{amssymb}

\usepackage[square,numbers]{natbib}
\usepackage{hyperref}

\usepackage{listings}

\usepackage{graphicx}
\graphicspath{{fig/}}

\usepackage{pdfpages}
\usepackage{float}

\usepackage{makecell}

\usepackage{url}

\usepackage{pgfplots}
\pgfplotsset{compat=newest}

\usepackage{algorithm}
\usepackage{algorithmic}

\usepackage{subfigure}

\usepackage[numbib]{tocbibind}

\usepackage{amsthm}

\usepackage{mathrsfs}
\usepackage{relsize}
\usepackage{exscale}

\newcommand{\realnumbers}{\mathbb{R}}
\newcommand{\nsd}{n_{\te{sd}}}
\newcommand{\ndof}{n_{\te{dof}}}

\newcommand{\vect}[1]{\boldsymbol{\mathrm{#1}}} 
\newcommand{\ten}[1]{\boldsymbol{\mathrm{#1}}} 

\newcommand{\operator}[1]{\mathscr{#1}}

\newcommand{\symgrad}[1]{\ten{\varepsilon}\br{#1}} 

\newcommand{\transpose}[1]{  {#1}^{\,\te{T}}  }

\newcommand{\br}[1]{\left ( #1 \right )} 
\newcommand{\brs}[1]{\left [ #1 \right ]} 

\newcommand{\ip}[2]{\left ( #1,#2 \right )} 
\newcommand{\ipe}[3]{\sum_{e=1}^{n_{\text{el}}} #1 \left ( #2,#3 \right )_e} 

\newcommand{\te}[1]{\text{#1}} 

\newcommand{\norm}[1]{\left\lVert#1\right\rVert} 
\newcommand{\abs}[1]{\left | #1\right |} 
\newcommand{\mydot}[0]{\cdot} 

\newcommand{\gradu}[0]{\nabla\vect{u}} 

\newcommand{\divu}[0]{\nabla\mydot\vect{u}} 
 
\newcommand{\divv}[0]{\nabla\mydot\vect{v}} 
 
\newcommand{\gradp}[0]{\nabla p} 
\newcommand{\epsu}[0]{ \ten{\varepsilon} \br{\vect{u}} } 
\newcommand{\epsv}[0]{ \ten{\varepsilon} \br{\vect{v}} } 

\newcommand{\etaT}[0]{\eta_{0}} 
\newcommand{\etaS}[0]{\eta_{\te{s}}} 
\newcommand{\etaP}[0]{\eta_{\te{p}}} 

\newcommand{\Wi}[0]{\te{Wi}} 

\newcommand{\cspace}[1]{\mathcal{#1}} 
\newcommand{\dspace}[1]{\mathcal{#1}_h} 

\begin{document}

\begin{titlepage}
	\begin{center}
		\vspace{0.5cm}
		\vspace{0.5cm}
		\hrule
		\vspace{3cm}
		\normalsize{Manuscript}\\
		\vspace{0.5cm}
		\LARGE{Stabilized finite element methods for a fully-implicit logarithmic reformulation of the Oldroyd-B constitutive law}\\
		\vspace{3cm}
		\Large{Stefan Wittschieber$^{a,*}$, Leszek Demkowicz$^b$, Marek Behr$^a$}\\
		\vspace{0.5cm}
		\normalsize{\today}\\
		\vspace{3cm}
		\begin{tabular}{l}
			$^*$ Corresponding author. \\ \\
			Email: \\
			wittschieber@aices.rwth-aachen.de (S. Wittschieber),  \\
			leszek@oden.utexas.edu (L. Demkowicz), \\
			behr@cats.rwth-aachen.de (M. Behr). \\ \\
			Affiliations:\\
			$^a$ Chair for Computational Analysis of Technical Systems, \\
			Center for Simulation and Data Science (JARA-CSD), \\
			RWTH Aachen University, Aachen, Germany \\ \\ 
			$^b$ Oden Institute for Computational Engineering and Sciences, \\
			The University of Texas at Austin, Austin, USA
		\end{tabular}		
	\end{center}
\end{titlepage}	

\section{Abstract}

Logarithmic conformation reformulations for viscoelastic constitutive laws have alleviated the high Weissenberg number problem, and the exploration of highly elastic flows became possible.
However, stabilized formulations for logarithmic conformation reformulations in the context of finite element methods have not yet been widely studied. We present stabilized formulations for the logarithmic reformulation of the Oldroyd-B model by Saramito (2014) based on the Variational Multiscale framework and Galerkin/Least-Squares method. The reformulation allows the use of Newton's method due to its fully-implicit nature for solving the steady-state problem directly. The proposed stabilization methods cure instabilities of convection and compatibility while preserving a three-field problem.
The spatial accuracy of the formulations is assessed with the four-roll periodic box, revealing comparable accuracy between the methods. The formulations are validated with benchmark flows past a cylinder and through a 4:1 contraction. We found an excellent agreement to benchmark results in the literature. The algebraic sub-grid scale method is highly robust, indicated by the high limiting Weissenberg numbers in the benchmark flows.\\

Keywords: Oldroyd-B, logarithmic reformulation, Stabilized finite element method, Sub-grid scale, 4:1 contraction, Flow past a cylinder

\section{Introduction}

In a variety of applications, the microstructure of the flowing medium causes viscoelastic behavior.
Examples of such fluids include molten plastics in industrial extrusion processing or blood in ventricular assist devices.
In addition to momentum and mass conservation laws, a hyperbolic constitutive equation provides a macroscopic description of viscoelasticity.
These equations pose several challenges for a numerical simulation. 
A unique challenge involves the computation of flow with increasing Weissenberg number, a non-dimensional measure relating elastic and viscous forces. 
Early attempts showed that computations with standard methods, such as the Galerkin finite element method, suffered from a lack of convergence in the iterative process at low Weissenberg numbers. This problem became known as high Weissenberg number problem (HWNP) \cite{Owens2002}.
It became evident that the conformation tensor, an intrinsic property of the constitutive equation, has to be symmetric positive definite (SPD) and that the loss of this property is connected to a breakdown in a computation \cite{Hulsen1990}. 
Developments in the 21st century have produced matrix-transformed conformation formulations. The original constitutive equation is replaced by an equivalent representation derived from a particular transformation of the conformation tensor that guarantees the SPD property. 
Besides a square root conformation reformulation by Balci \cite{Balci2011}, the logarithmic conformation (log-conf) reformulation by Fattal and Kupferman \cite{Fattal2004} and its variations are the most frequent choices in the development of new numerical methods.
Subsequent works confirmed that these reformulations improve the stability of numerical computations of viscoelastic flows significantly, i.e., flows at higher Weissenberg numbers became possible to simulate \cite{Hulsen2005}, \cite{Pimenta2017}, \cite{Afonso2011}.

Previous works of Knechtges \cite{Knechtges2014a} and Saramito \cite{Saramito2014} developed fully-implicit logarithmic conformation formulations, which can be stated in a closed form together with the equations for momentum and mass. The Newton-Raphson method can be applied directly to the steady-state problem, as no iterative coupling is necessary, as opposed to the original log-conf formulation by Fattal and Kupfermann.
Another benefit of Saramito's formulation is that it does not degenerate at zero Weissenberg number. Thus, continuation techniques in the Weissenberg number can be applied for steady-state computations starting with the Newtonian case. 

Besides the HWNP, the three-field problem --- with the unknowns pressure, velocity, and viscoelastic stresses --- suffers from two other sources of instabilities in a finite-element context.
Standard finite elements represent the convective terms of the momentum and constitutive equation inadequately. The other instability may emanate from an inappropriate choice of interpolation spaces. Compatibility conditions between pressure-velocity interpolation and velocity-stress interpolation have to be satisfied in the discrete problem.

In the recent developments of numerical methods for viscoelastic flows, Streamline-Upwind Petrov-Galerkin (SUPG) methods \cite{Marchal87a} and Discontinuous Galerkin (DG) methods \cite{Fortin1989} are still the main tools to overcome problems from the convective terms.
To avoid the design of stress elements that fulfill the compatibility, the SUPG methods can be combined with the Discrete Elastic Viscous Stress Splitting (DEVSS) method \cite{Guenette1995}, where the velocity gradient is treated as an unknown, e.g., \cite{Varchanis2019}.
Due to the high number of degrees of freedom in the descriptive equations of viscoelasticity, one strives for methods that can be applied without introducing other unknowns to reduce computational efforts.

The Galerkin Least-Squares (GLS) method, and methods evolving from the Variational Multiscale (VMS) framework, introduce a stream-line diffusion term and a DEVSS-like term without introducing new unknowns.
Both methods were applied to standard formulations in the context of viscoelastic constitutive laws. Coronado \cite{Coronado2005a} proposed a GLS method for the four-field formulation, and Kwack \cite{Kwack2010} and Castillo \cite{Castillo2014} developed stabilized formulations in the VMS framework.

However, the development of these stabilized formulations for log-conf reformulations is limited and is still an area of current research.
Moreno \cite{Moreno2019} applied a VMS approach with orthogonal sub-scales to the log-conf formulation by Coronado \cite{Coronado2007} with a modification to remove the degeneracy in the Newtonian limit. In this work, we combine the unique benefits of the non-singular log-conf formulation for the Oldroyd-B model by Saramito \cite{Saramito2014} and a stabilized formulation based on the VMS framework to obtain a robust, computationally efficient alternative to existing approaches. The sub-grid scales are assumed to be proportional to the finite element residuals. The method will be referred to as the algebraic sub-grid scale (ASGS) method \cite{Castillo2014} and \cite{Moreno2019}. Since the VMS framework relies on a specific linearized representation of the PDE system, we propose a SUPG and GLS method that evolved during the design process.

The article is structured as follows: Sec.~\ref{sec:viscFlowProb} presents the governing equations, followed by a short review of the log-conf formulation. The variational problem is stated, and from it, the Galerkin discretization is derived. Sec.~\ref{sec:designStabMethods} presents our stabilized methods, including a proposal for stabilization parameters. Sec.~\ref{sec:numResults} presents the numerical results for the four-roll periodic box, and two benchmark flows past a cylinder and through a planar 4:1 contraction. Sec.~\ref{sec:conclusion} contains the conclusions.

\section{The viscoelastic flow problem}
\label{sec:viscFlowProb}

\subsection{Governing equations}
\label{sec:governingeqn}

In this work, an incompressible and isothermal viscoelastic flow at steady state is considered. A fluid occupies a spatial domain $\Omega$ with its boundary $\Gamma$. The governing equations involve the conservation laws for momentum and mass:
\begin{align}
\rho \vect{u} \mydot \nabla \vect{u} - \nabla \mydot \ten{T} = \vect{b} & \quad \te{in} \quad \Omega, \label{eqn:momentum}\\
\nabla \mydot \vect{u} = 0 & \quad\te{in}\quad \Omega \label{eqn:mass},
\end{align}
where $\rho$ describes the constant density, $\vect{u}$ describes the velocity field, $\vect{b}$ is the volumetric force field, and $\ten{T}$ is the Cauchy stress tensor. It can be decomposed into a viscous and a viscoelastic contribution,
\begin{align}
\ten{T} = -p \ten{I} + 2 \etaS \epsu + \ten{\sigma} \,, \label{eqn:cauchy}
\end{align}
with the pressure field $p$, the solvent (or effective) viscosity $\eta_{\te{s}}$, the rate-of-strain tensor $\epsu = \frac{1}{2} \br{ \gradu + \transpose{\gradu}}$, the velocity gradient $\gradu := \dfrac{\partial u_i}{\partial x_j}\vect{e}_i\vect{e}_j$ in Cartesian basis $\vect{e}_i$, and the extra stresses (or viscoelastic stresses) $\ten{\sigma}$. In order to close the system of equations, a constitutive relation has to be modeled. In this work, the differential Oldroyd-B model \cite{Oldroyd1950} is considered,
\begin{align}
\lambda \br{ \vect{u} \mydot \nabla \ten{\sigma} - \nabla \vect{u} \mydot \ten{\sigma} - \ten{\sigma} \mydot \transpose{\br{\nabla \vect{u}}}} + \ten{\sigma} = 2 \etaP \epsu\quad \te{in} \quad \Omega\,, \label{eqn:constsig}
\end{align}
where $\lambda$ is the relaxation time, and $\etaP$ is the polymeric viscosity. The first term on the left-hand side of Eqn.~(\ref{eqn:constsig}) is the upper-convected tensor derivative of the extra stress tensor. Boundary conditions for the velocity have to be imposed on all boundaries. Due to the hyperbolic nature of the constitutive equation, the extra stresses have to be prescribed solely on the inflow part of the boundary, where $\vect{u}\mydot\vect{n}<0$ with outward unit normal $\vect{n}$.

\subsection{The log-conf reformulation}
\label{sec:logconfformulation}

The constitutive equation (\ref{eqn:constsig}) can be written in terms of the conformation tensor $\ten{c} := \ten{\sigma} + \frac{\etaP}{\lambda} \ten{I}$:
\begin{align}
\lambda \left ( \vect{u}\mydot\nabla\ten{c} - \nabla\vect{u} \mydot\ten{c} - \ten{c}\mydot\transpose{\br{\nabla\ten{u}}} \right)+ \ten{c} = \frac{\etaP}{\lambda} \ten{I}\,.
\label{eqn:constc}
\end{align} 
This equation is the starting point for most derivations of logarithmic reformulations. The standard approach is to derive an equation from Eqn.~(\ref{eqn:constc}) for a new variable $\ten{\chi}$ such that $\ten{c} = \exp{\br{\frac{\lambda}{\etaP}\ten{\chi}}}$ \cite{Fattal2004}, \cite{Coronado2007}, \cite{Knechtges2014a}.
The positive definiteness of the conformation tensor is, therefore, enforced.
Saramito \cite{Saramito2014} proposed a slight modification to the transformation,
\begin{align}
\ten{\chi} = \frac{\etaP}{\lambda} \log{\br{\frac{\lambda}{\etaP}\ten{c}}} \quad \Longleftrightarrow \quad\ten{c} = \frac{\etaP}{\lambda} \exp{\br{\frac{\lambda}{\etaP}\ten{\chi}}} \,.
\label{eqn:transformation}
\end{align}
The logarithmic conformation tensor $\ten{\chi}$ is given back the units of the (original) extra stresses, which yields a transformation that does not degenerate for $\lambda=0$.
To obtain the log-conf formulation, Eqn.~(\ref{eqn:constc}) is rewritten in the eigenbasis of $\ten{c}$, the eigenvalues are transformed by Eqn.~(\ref{eqn:transformation}), and the obtained relation is written back into Cartesian basis. This results in the equation
\begin{align}
&\lambda \left ( \vect{u}\mydot\nabla\ten{\chi} - \ten{\Omega} \br{\vect{u}} \mydot\ten{\chi} + \ten{\chi}\mydot\ten{\Omega}\br{\vect{u}} \right) \nonumber \\
&+ \ten{\chi} - \ten{f}\br{\frac{\lambda}{\etaP},\,-\ten{\chi}} + 2 \etaP \ten{\kappa} \br{\frac{\lambda}{\etaP}\ten{\chi},\, \symgrad{\vect{u}}} = 2 \eta_{\text{p}} \epsu\,,
\label{eqn:constchi}
\end{align}
where $\ten{\Omega}\br{\vect{u}} := \frac{1}{2}\br{\nabla \vect{u} - \transpose{\br{\nabla \vect{u}}}}$.

Both functions, $\ten{f}(\mydot,\mydot)$ and $\ten{\kappa}(\mydot,\mydot)$, satisfy $\ten{f}(0,\,\mydot)=\ten{0}$ and $\ten{\kappa}\br{\ten{0},\,\mydot}=\ten{0}$, which is why Eqn.~(\ref{eqn:constchi}) reduces to the Newtonian case continuously. In addition, both functions, are continuously differentiable everywhere, enabling Newton's method. The non-linear function $\ten{f}(\mydot,\,\mydot)$ relates the extra stresses $\ten{\sigma}$ to the logarithmic conformation tensor $\ten{\chi}$,
\begin{align}
\ten{\sigma} = \ten{\chi} + \ten{f}\br{\frac{\lambda}{\etaP},\,\ten{\chi}} \,.
\label{eqn:transformsigchi}
\end{align}
The definitions of $\ten{f}(\mydot,\,\mydot)$ and $\ten{\kappa}(\mydot,\,\mydot)$ are found in the original work \cite{Saramito2014}. In the following, we adhere to the original notation regarding the arguments of the functions.

For further discussions, we write Eqns.~(\ref{eqn:momentum}), (\ref{eqn:mass}), (\ref{eqn:cauchy}), (\ref{eqn:constchi}), and  (\ref{eqn:transformsigchi}) compactly as
\begin{align}
\operator{L}(\vect{u};\,U) = F \,, \label{eqn:abstractProblem}
\end{align}  
where $U:=\brs{\vect{u},\,p,\,\ten{\chi}}$, $F:=\brs{\vect{b},\,0,\,\ten{0}}$, 
\begin{align}
\operator{L}(\vect{u}^*;U) := \begin{pmatrix}
\rho \vect{u}^*\mydot\gradu - 2\beta\eta_0 \nabla \mydot \epsu - \nabla \mydot \ten{\chi} - \nabla \mydot \ten{f}\br{\frac{\lambda}{\etaP},\,\ten{\chi}} + \gradp \\
\divu\\
\frac{1}{2\eta_0}\br{\ten{\chi} - \ten{f}\br{\frac{\lambda}{\etaP},\,-\ten{\chi}}} - \br{1-\beta}\epsu \\
\quad + \frac{\lambda}{2\eta_0}\br{\vect{u}^* \mydot \nabla \ten{\chi} + \ten{\chi} \mydot \ten{\Omega}\br{\vect{u}^*} - \ten{\Omega}\br{\vect{u}^*} \mydot \ten{\chi}} + \br{1-\beta} \ten{\kappa} \br{\frac{\lambda}{\etaP}\ten{\chi},\, \symgrad{\vect{u}^*}} \end{pmatrix} \,,
\label{eqn:diffoperator}
\end{align}
with the total (or apparent) viscosity $\etaT = \etaS + \etaP$, and the solvent viscosity ratio $\beta=\etaS / \etaT$.
Note that we have introduced the advective velocity $\vect{u}^*$ to highlight the non-linearities in the convective terms.
As $\ten{\kappa}(\mydot,\mydot)$ is linear in its second argument, the operator $\operator{L}(\vect{u}^*;U)$ is, in fact, linear in the velocity $\vect{u}$. 

\subsection{Variational problem and Galerkin discretization}
\label{sec:varprob}

In this section, the variational problem is stated. For this purpose, the following spaces on domain $\Omega$ are introduced: the space of square integrable functions is denoted by $L^2(\Omega)$, and the space in which every function whose first distributional derivative belongs to $L^2(\Omega)$ is denoted by $H^1(\Omega)$. The $L^2$ inner product is denoted by $\ip{\cdot}{\cdot}$ for scalars, vectors, and tensors. For two tensors $\ten{A}$ and $\ten{B}$, it is defined as $\ip{\ten{A}}{\ten{B}}:=\int_{\Omega} \ten{A}:\ten{B}\,\te{d}\Omega$. Let the function spaces for velocity, pressure, and elastic stresses be $\cspace{V}:=H^1(\Omega)^{\nsd}$, $\cspace{Q}:=L^2(\Omega)/\realnumbers$, and $\cspace{T}:=H^1(\Omega)^{\nsd\times\nsd}_{\te{sym}}$, respectively. $\nsd$ is the number of space dimensions. The variational formulation for (\ref{eqn:abstractProblem}) reads: find $U = \brs{\vect{u},\,p,\,\ten{\chi}} \in \cspace{S} := \cspace{V} \times \cspace{Q} \times \cspace{T}$, such that
\begin{align}
B(\vect{u};U,V) = L(V) := \ip{\vect{b}}{\vect{v}}
\label{eqn:varprob}
\end{align}
for all $V = \brs{\vect{v},\,q,\,\ten{\tau}} \in \cspace{S}$, with
\begin{align}
B&(\vect{u}^*;U,V) = 2 \beta \eta_0 \ip{\epsu}{\epsv} + \ip{\rho\, \vect{u}^* \mydot \gradu}{\vect{v}} - \ip{p}{\divv} \nonumber \\ 
& + \ip{\ten{\chi} + \ten{f}\br{\frac{\lambda}{\etaP},\,\ten{\chi}}}{\epsv}  +\ip{\divu}{q} \nonumber\\
&+ \frac{1}{2 \eta_0} \ip{\ten{\chi} - \ten{f}\br{\frac{\lambda}{\etaP},\,-\ten{\chi}}}{\ten{\tau}} - \br{1-\beta} \ip{\epsu}{\ten{\tau}} \nonumber \\ 
& + \frac{\lambda}{2 \eta_0}\ip{\vect{u}^* \mydot \nabla \ten{\chi} + \ten{\chi} \mydot \ten{\Omega}\br{\vect{u}^*} - \ten{\Omega}\br{\vect{u}^*} \mydot \ten{\chi}}{\ten{\tau}} \nonumber \\ 
&+ \br{1-\beta} \ip{\ten{\kappa} \br{\frac{\lambda}{\etaP}\ten{\chi},\, \symgrad{\vect{u}^*}}}{\ten{\tau}}\,,
\label{eqn:nonlinearform}
\end{align}
where integration by parts was performed on the Cauchy stress tensor.

Given an admissible tessellation $\Omega_h$, conforming finite element spaces $\dspace{V}\subset\cspace{V}$, $\dspace{Q}\subset\cspace{Q}$ and $\dspace{T}\subset\cspace{T}$ are constructed. The Galerkin finite element discretization reads: find $U_h = \brs{\vect{u}_h,\,p_h,\,\ten{\chi}_h} \in \dspace{S} := \dspace{V} \times \dspace{Q} \times \dspace{T}$ such that
\begin{align}
B(\vect{u}_h;U_h,V_h) = L(V_h)
\end{align}
for all $V_h = \brs{\vect{v}_h,\,q_h,\,\ten{\tau}_h} \in \dspace{S}$.

Throughout this work, the same trial and test approximation spaces are used. Linear or bilinear first-order elements are used for all fields for triangular or quadrilateral meshes, respectively.

\section{Design of stabilized methods}
\label{sec:designStabMethods}

\subsection{Linearized operator}
\label{sec:Linearization}

The design of VMS schemes relies on a specific linearization of the original non-linear operator (\ref{eqn:diffoperator}).
The linearization is computed by the directional derivative
\begin{align}
& \operator{L}^{\te{lin}}(\vect{u}^*,\ten{\chi}^*;\delta U) := \frac{\te{d}}{\te{d}\varepsilon}\operator{L}(\vect{u}^*;U^*+\varepsilon\delta U) \Big|_{\varepsilon=0} \nonumber \\
& = \begin{pmatrix}
\rho \vect{u}^*\mydot\nabla \delta \vect{u} - 2\beta\eta_0 \nabla \mydot \symgrad{\delta \vect{u}} - \nabla \mydot \delta\ten{\chi} - \nabla \mydot \te{D}\ten{f}\br{\frac{\lambda}{\etaP},\,\ten{\chi}^*}\brs{\delta\ten{\chi}} + \nabla \delta p \\
\nabla \mydot \delta \vect{u}\\
\frac{1}{2\eta_0}\br{\delta \ten{\chi} + \te{D}\ten{f}\br{\frac{\lambda}{\etaP},\,-\ten{\chi}^*}}\brs{\delta \ten{\chi}} - \br{1-\beta}\symgrad{\delta \vect{u}} \\
\quad + \frac{\lambda}{2\eta_0}\br{\vect{u}^* \mydot \nabla \delta \ten{\chi} + \delta \ten{\chi} \mydot \ten{\Omega}\br{\vect{u}^*} - \ten{\Omega}\br{\vect{u}^*} \mydot \delta \ten{\chi}} + \frac{\lambda}{\etaT} \te{D}\ten{\kappa} \br{\frac{\lambda}{\etaP}\ten{\chi}^*,\, \symgrad{\vect{u}^*}}\brs{\delta \ten{\chi}} \end{pmatrix} \,.
\label{eqn:lindiffoperator}
\end{align}
Note that the advective velocities are treated as a known field. The advective velocities will be later replaced with the unknown velocity field. This choice is a simplification to reduce the complexity of the numerical scheme, see also \cite{Coronado2005a} or \cite{Castillo2014}. The capital $\te{D}$ denotes the Gateaux derivative, and the square brackets $\brs{\mydot}$ denote the direction.
Algebraic expressions for the derivatives of $\ten{f}(\mydot,\mydot)$ and $\ten{\kappa}(\mydot,\mydot)$, were obtained by using the computer algebra system Maxima \cite{Maxima}.

\subsection{SUPG and GLS method}
\label{sec:gls}

Whereas in the standard formulation (\ref{eqn:constsig}), fixing the convective velocities produces a linearization, in the log-conf formulation it 
does not. The GLS method is based on the linearized presentation (\ref{eqn:lindiffoperator}) and reads: find $U_h \in \dspace{X}$, such that
\begin{align}
B(\vect{u}_h;U_h,V_h) + \ipe{\alpha}{\operator{L}(\vect{u}_h;U_h)-F}{\operator{L}^{\te{lin}}(\vect{u}_h,\ten{\chi}_h;V_h)} = L(V_h)\label{eqn:GLS}
\end{align}
for all $V_h \in \dspace{X}$, with the element-wise inner product $\ip{\mydot}{\mydot}_e$, and $n_{\text{el}}$ being the number of elements of the tessellation. 

$\alpha$ is the matrix of stabilization parameters. We assume the stabilization matrix of diagonal form $\alpha = \te{diag}\br{\alpha_{\te{mom}},\alpha_{\te{con}},\alpha_{\te{const}}}$. Note, we have re-introduced the non-linearity in the convective terms by taking $\vect{u}^*=\vect{u}_h$. Depending on how the method is embedded into a non-linear iteration strategy, other choices are possible, e.g., taking $\vect{u}^*$ directly as velocity field of the previous iteration \cite{Castillo2014}. The stabilized method (\ref{eqn:GLS}) introduces further non-linearities to the trial functions, not only due to the latter decision but also due to the dependence of the stabilization parameters on the unknowns $U$.

We define the finite element residuals as
\begin{align}
\vect{R}_h^{\te{mom}} & = \rho \vect{u}_h\mydot\gradu_h - \nabla \mydot \ten{\chi}_h - \nabla \mydot \ten{f}\br{\frac{\lambda}{\etaP},\,\ten{\chi}_h}  + \gradp_h - \vect{b}_h \\
R_h^{\te{con}} &= \nabla \mydot \vect{u}_h \\ 
\ten{R}_h^{\te{const}} &= \frac{1}{2\eta_0}\br{\vect{\chi}_h - \ten{f}\br{\frac{\lambda}{\etaP},\,-\ten{\chi}_h}} - \br{1-\beta}\symgrad{\vect{u}_h} \nonumber\\
&+ \frac{\lambda}{2 \eta_0}\br{\vect{u}_h \mydot \nabla \ten{\chi}_h + \ten{\chi}_h \mydot \ten{\Omega}\br{\vect{u}_h} - \ten{\Omega}\br{\vect{u}_h} \mydot \ten{\chi}_h} \nonumber\\
& + \frac{\lambda}{\eta_0} \ten{\kappa} \br{\frac{\lambda}{\etaP}\ten{\chi}_h,\, \symgrad{\vect{u}_h}} \,.
\end{align}
Now, we can write the stabilized method (\ref{eqn:GLS}) in the following way: find $U_h \in \dspace{X}$, such that
\begin{align}
&B(\vect{u}_h;U_h,V_h) + \sum_{e=1}^{n_{\text{el}}} \alpha_{\te{con}} \left ( R_h^{\te{con}} \, ,\divv_h \right )_{\te{e}}
+ \sum_{e=1}^{n_{\text{el}}} \alpha_{\te{mom}} \Big ( \vect{R}_h^{\te{mom}} , - \nabla \mydot \ten{\tau}_h + \rho \vect{u}_h \mydot  \nabla \vect{v}_h + \nabla q_h  \Big )_{\te{e}} \nonumber \\
& + \sum_{e=1}^{n_{\text{el}}} \alpha_{\te{const}} \Big ( \ten{R}_h^{\te{const}} , \operator{P} \br{\vect{u}_h,\ten{\chi}_h; V_h} \Big )_{\te{e}} = L(V_h)
\label{eqn:glssupg}
\end{align}
for all $V_h \in \dspace{X}$. In the following, we refer to the SUPG method, or GLS method, by taking $\operator{P} \br{\vect{u}^*,\ten{\chi}^*; V}$ either
\begin{align}
\operator{P}^{\te{SUPG}} \br{\vect{u}^*,\ten{\chi}^*; V} &= \frac{\lambda}{2 \eta_0} \vect{u}^* \mydot \nabla \ten{\tau},\,\te{or} \label{eqn:projsupg}\\
\operator{P}^{\te{GLS}} \br{\vect{u}^*,\ten{\chi}^*; V} &= \frac{1}{2 \eta_0} \br{\ten{\tau} + \te{D}\ten{f}\br{\frac{\lambda}{\etaP},\,-\ten{\chi}^*}\brs{\ten{\tau}}}  - \br{1-\beta} \symgrad{\vect{v}} \nonumber \\ 
&+ \frac{\lambda}{2 \eta_0}\br{\vect{u}^* \mydot \nabla \ten{\tau} + \ten{\tau} \mydot \ten{\Omega}\br{\vect{u}^*} - \ten{\Omega}\br{\vect{u}^*} \mydot \ten{\tau}} + \frac{\lambda}{\eta_0} \te{D}\ten{\kappa} \br{\frac{\lambda}{\etaP}\ten{\chi}^*,\, \symgrad{\vect{u}^*}}\brs{\ten{\tau}}\,. \label{eqn:projgls}
\end{align}

We deleted the divergence of the strain $2\beta\eta_0 \nabla \mydot \symgrad{u_h}$, as second derivatives of linear elements either vanish or become insignificant in the case of bilinear elements. Hence, the method is only consistent in a weak sense, as highlighted in \cite{Jansen1999}.
In addition, we deleted also the non-linearity $\te{D}\ten{f}\br{\frac{\lambda}{\etaP},\,\ten{\chi}^*}\brs{\ten{\tau}}$ in the stabilization of the momentum equation. Based on our experience, the inclusion of this term was harmful to the robustness of the method.

\subsection{Algebraic sub-grid scale method}
\label{sec:asgs}

The VMS framework can only be applied to a certain linearization for non-linear problems. Rather than relying on a linearization of the standard formulation \cite{Moreno2019}, we construct our method on the linearization (\ref{eqn:lindiffoperator}) obtained from the log-conf formulation. 
The first step is to split the unknown $U$ into the finite element solution $U_h$, and the fine-scale solution $\tilde{U}$ \cite{Hughes1995}. Introducing the decomposition, and integrating by parts yields the finite element problem: given $\vect{u}^*$ and $\ten{\chi}^*$, find $U_h \in \dspace{X}$ such that
\begin{align}
B^{\te{lin}}(\vect{u}^*,\ten{\chi}^*;U_h,V_h) + \ipe{}{\operator{L}^{\te{lin}}(\vect{u}^*;\ten{\chi}^*;\tilde{U})}{V_h} = L(V_h) \label{eqn:asgs}
\end{align}
for all $V_h \in \dspace{X}$. Here, $B^{\te{lin}}(\vect{u}^*,\ten{\chi}^*;U_h,V_h)$ denotes the bilinear form obtained from linearizing the form (\ref{eqn:nonlinearform}) following the same steps as in Sec.~\ref{sec:Linearization}. We introduced this form for formal reasons. Practically, this linearization can only be embedded into a fixed-point iteration. For the final method, the non-linear form $B(\vect{u}^*;U,V)$ will be used.
Next, integration by parts is performed on those contributions of the second term of (\ref{eqn:asgs}) that contain derivatives on the fine-scales:
\begin{align}
& \ipe{}{\operator{L}^{\te{lin}}(\vect{u}_h^*;\ten{\chi}_h^*;\tilde{U})}{V_h} \nonumber \\
& = - \ipe{}{\tilde{\vect{u}}}{\rho \vect{u}^*_h\mydot\nabla \vect{v}_h} + \ipe{}{\tilde{\ten{\chi}} + \te{D}\ten{f}\br{\frac{\lambda}{\etaP},\,\ten{\chi}^*_h}\brs{\tilde{\ten{\chi}}}}{\symgrad{\vect{v}_h}} \nonumber \\
& - \ipe{}{\tilde{\vect{u}}}{2\beta\eta_0 \nabla \mydot \symgrad{\vect{v}_h} } - \ipe{}{\tilde{p}}{\nabla \mydot \vect{v}_h} - \ipe{}{\tilde{\vect{u}}}{\nabla q_h} \nonumber \\
& - \ipe{}{\tilde{\ten{\chi}}}{\frac{\lambda}{2 \eta_0} \vect{u}^*_h \mydot \nabla \ten{\tau}_h } + \ipe{}{\frac{\lambda}{2 \eta_0}\br{\tilde{\ten{\chi}} \mydot \ten{\Omega}\br{\vect{u}^*_h} - \ten{\Omega}\br{\vect{u}^*_h} \mydot \tilde{\ten{\chi}}}}{\ten{\tau}_h} \nonumber \\
& + \ipe{}{\frac{1}{2\eta_0}\br{\tilde{\ten{\chi}} + \te{D}\ten{f}\br{\frac{\lambda}{\etaP},\,-\ten{\chi}^*_h}\brs{\tilde{\ten{\chi}}}}}{\ten{\tau}_h} + \ipe{}{\frac{\lambda}{\eta_0} \te{D}\ten{\kappa} \br{\frac{\lambda}{\etaP}\ten{\chi}^*_h,\, \symgrad{\vect{u}^*_h}}\brs{\tilde{\ten{\chi}}}}{\ten{\tau}_h} \nonumber \\
& + \ipe{}{\tilde{\vect{u}}}{\br{1-\beta}\nabla \mydot \ten{\tau}_h}\,,
\end{align}
without considering boundary conditions. The idea behind this last step is that it is only possible to capture the integral effects of the fine scales, whereas it is difficult to approximate their derivatives. The algebraic sub-grid scale method is obtained by taking the fine scales proportional to the finite element residuals \cite{Castillo2014}, i.e.,
\begin{align}
\tilde{U} = \begin{pmatrix} \tilde{\vect{u}}\\ \tilde{p}\\ \tilde{\ten{\chi}} \end{pmatrix} \approx \begin{pmatrix} - \alpha_{\te{mom}} \vect{R}_h^{\te{mom}}\\ - \alpha_{\te{con}} R_h^{\te{con}}\\ - \alpha_{\te{const}} \ten{R}_h^{\te{const}} \end{pmatrix} \,.
\end{align}
After further rearrangements, our ultimate method is obtained: find $U_h \in \dspace{X}$ such that
\begin{align}
&B(\vect{u}_h;U_h,V_h) + \ipe{\alpha_{\te{mom}}}{\vect{R}_h^{\te{mom}}}{\rho \vect{u}_h\mydot\nabla \vect{v}_h + \nabla q_h - \br{1-\beta}\nabla \mydot \ten{\tau}_h} \nonumber \\
& - \ipe{\alpha_{\te{const}}}{\ten{R}_h^{\te{const}} + \te{D}\ten{f}\br{\frac{\lambda}{\etaP},\,\ten{\chi}_h}\brs{\ten{R}_h^{\te{const}}}}{\symgrad{\vect{v}_h}} \nonumber \\
& + \ipe{\alpha_{\te{con}}}{\nabla \mydot \vect{u}_h}{\nabla \mydot \vect{v}_h} \nonumber \\
& + \ipe{\alpha_{\te{const}}}{\ten{R}_h^{\te{const}}}{\frac{\lambda}{2 \eta_0} \vect{u}_h \mydot \nabla \ten{\tau}_h } - \ipe{\alpha_{\te{const}}}{\frac{\lambda}{2 \eta_0}\br{\ten{R}_h^{\te{const}} \mydot \ten{\Omega}\br{\vect{u}_h} - \ten{\Omega}\br{\vect{u}_h} \mydot \ten{R}_h^{\te{const}}}}{\ten{\tau}_h} \nonumber \\
& - \ipe{\alpha_{\te{const}}}{\frac{1}{2\eta_0}\br{\ten{R}_h^{\te{const}} + \te{D}\ten{f}\br{\frac{\lambda}{\etaP},\,-\ten{\chi}_h}\brs{\ten{R}_h^{\te{const}}}}}{\ten{\tau}_h} \nonumber \\
& - \ipe{\alpha_{\te{const}}}{\frac{\lambda}{\eta_0} \te{D}\ten{\kappa} \br{\frac{\lambda}{\etaP}\ten{\chi}_h,\, \symgrad{\vect{u}_h}}\brs{\ten{R}_h^{\te{const}}}}{\ten{\tau}_h} \nonumber \\
&= L(V_h)
\label{eqn:asgsfinal}
\end{align}
for all $V_h \in \dspace{X}$. Again, the divergence of the strain rate was deleted. If we compare the ASGS method (\ref{eqn:asgsfinal}) to the GLS method (\ref{eqn:glssupg}),(\ref{eqn:projgls}), we see that the stabilization for the momentum equation is nearly unchanged. The term in the second line in (\ref{eqn:asgsfinal}) contains, among other things, an elliptic term that gives control on the velocity. The term in the third line stabilizes at high Reynolds numbers. It can be neglected here but is left for generality. The first term in the fourth line stabilizes the convection of the stresses and coincides with the SUPG stabilization term in (\ref{eqn:projsupg}).


\subsection{Stabilization parameters}
\label{sec:stabParam}

The choice of the stabilization parameters $\alpha_{\te{mom}}$, $\alpha_{\te{con}}$, and $\alpha_{\te{const}}$ is essential in the design of every stabilized method. For the design of $\alpha_{\te{mom}}$ and $\alpha_{\te{con}}$, we adapt the definitions of \cite{Bazilev2007}:
\begin{align}
\alpha_{\te{mom}} &= \brs{\etaT \sqrt{\ten{G}\mydot\mydot\ten{G}} + \rho \sqrt{\vect{u}_h\mydot\ten{G}\vect{u}_h} }^{-1} \\
\alpha_{\te{con}} &= \brs{\alpha_{\te{mom}} \te{tr}\br{\ten{G}}}^{-1}
\end{align}
The co-variant metric tensor $\ten{G}$, with its components $G_{ij}=\frac{\partial \xi_k}{\partial x_i}\frac{\partial \xi_k}{\partial x_j}$, allows a convenient construction of a directional element length. $\frac{\partial \vect{\xi}}{\partial \vect{x}}$ is the inverse of the Jacobian of the mapping from the reference element coordinates to the global coordinates.

For the SUPG method, we adapt the definition proposed in \cite{Knechtges2014a}:
\begin{align}
\alpha^{\te{SUPG}}_{\te{const}} &= \brs{ \frac{1}{2\etaT} + \frac{\lambda}{2\etaT} \sqrt{\vect{u}_h\mydot\ten{G}\vect{u}_h} }^{-1}
\end{align}
The purpose of the contribution $\frac{1}{2\etaT}$ is to still obtain a well-behaved value in the limit, where either $\lambda\rightarrow0$, or $\vect{u}_h\rightarrow\vect{0}$.

Several numerical experiments suggested the following choice for $\alpha_{\te{const}}$ for the GLS and ASGS method:
\begin{align}
&\alpha_{\te{const}} =\Bigg [ \frac{\lambda}{2\etaT} \sqrt{\vect{u}_h\mydot\ten{G}\vect{u}_h} + \frac{\lambda}{\etaT} \norm{\ten{\Omega}\br{\vect{u}_h}}_2 \nonumber \\
&+ \frac{1}{2\etaT} \sum_{m=1}^{M} \norm{\ten{E}_m + \frac{\partial \ten{f}}{\partial \zeta_m}\br{\frac{\lambda}{\etaP},\,-\ten{\chi}_h}}_2 + \frac{\lambda}{\etaT} \sum_{m=1}^{M} \norm{\frac{\partial\ten{\kappa}}{\partial \beta_m}\br{\frac{\lambda}{\etaP}\ten{\chi}_h,\, \symgrad{\vect{u}_h}}}_2 \Bigg ]^{-1} \,.
\end{align}
$\norm{\mydot}_2$ denotes the Frobenius norm, $M = \nsd \br{\nsd +1} / 2$, and $\ten{E}_m$ denotes the standard basis for symmetric matrices.  $\beta_m$ and $\zeta_m$ refer to the linearly independent components of the functional arguments $\ten{\beta}$ and $\ten{\zeta}$ of the original work \cite{Saramito2014}, respectively. The inclusion of the non-linearities, $\ten{f}(\mydot,\mydot)$ and $\ten{\kappa}(\mydot,\mydot)$ in the stabilization parameter, appeared to be crucial to obtain convergence in highly elastic flows. All stabilization parameters are computed locally for every quadrature point of each element.

\subsection{Linearization of the stabilized problem}
\label{sec:LinFD}

The Newton method is used to solve the stabilized variational problems (\ref{eqn:glssupg}), and (\ref{eqn:asgsfinal}).
The linear problem to be solved for the new incremental step $\delta U_h^i$ in each non-linear iteration step $i$ reads
\begin{align}
\te{D} \ten{Res}\br{U_h^{i-1}}\brs{\delta U_h^i} = - \ten{Res}\br{U_h^{i-1}}\,,
\label{eqn:linprob}
\end{align}
where we denote the discrete residuals of (\ref{eqn:glssupg}) or (\ref{eqn:asgsfinal}) compactly as $\ten{Res}(U_h)$. The directional derivative may be also written in terms of the Jacobian matrix $\frac{\partial \ten{Res}}{\partial U_n}$, i.e.,
\begin{align}
\te{D} \ten{Res}\br{U^*}\brs{\delta U} = \sum_{n=1}^{\ndof} \delta U_n \frac{\partial \ten{Res}}{\partial U_n}\br{U^*} \,,
\end{align}
with $U_n$ referring to the components of $U$. We use a central finite difference stencil to compute the Jacobians:
\begin{align}
\frac{\partial \ten{Res}}{\partial U_n}\br{U^*} \approx \frac{\ten{Res}\br{U_n^* + \Delta_{\te{FD}}} - \ten{Res}\br{U_n^* - \Delta_{\te{FD}}}}{2 \Delta_{\te{FD}}} \,.
\label{eqn:centralfdscheme}
\end{align}
A complete analytic approach would be preferable to reduce computational effort. Whereas $2\ndof$ evaluations of the residuals are required in two spatial dimensions for central finite differences, only a single routine call is required for an analytic Jacobian.
Nonetheless, obtaining an analytic Jacobian is highly complex due to the additional non-linearities in the stabilization.
This aspect is discussed further in Sec.~(\ref{sec:anavsfd}). We take the step size $\Delta_{\te{FD}}=10^{-5}$ throughout this work.

The non-linear iteration loop is enclosed in a continuation procedure in the relaxation time to obtain converged solutions at highly elastic flows. A computation is considered converged, if $\norm{\ten{Res}\br{U_h^{i}}} / \sqrt{\ndof}<\epsilon$, where $\norm{\mydot}$ is the Euclidean norm, and $\epsilon$ a user-defined threshold. The linear solver we employ is the FGMRES method \cite{Saad1993} with PARDISO \cite{Schenk2004} as right preconditioner. 

\section{Numerical results}
\label{sec:numResults}

\subsection{Four-roll periodic box}
\label{sec:fourroll}

We begin with a numerical study to assess the spatial order of accuracy of our methods. The test problem is characterized by a square domain, $\Omega = \brs{-L,L}\times\brs{-L,L}$, with periodicity conditions on all boundaries, and a body force field,
\begin{align}
\vect{b} = 2 F_s \begin{pmatrix} \sin(x) \cos(y) \\ -\cos(x) \sin(y)  \end{pmatrix} \,
\end{align}
that generates four vortices, causing a contraction of the fluid along the $y$-axis and stretching along the $x$-axis. $F_s$ is a scaling constant taken as one. The Weissenberg number is defined as $\Wi = \lambda \frac{L F_s}{\pi \beta \etaT}$. We take $\beta = 2/3$, $\etaT=1$, $L=\pi$, and assume creeping flow conditions. This flow configuration was studied in a particular view of the Oldroyd-B model \cite{Thomases2007}. The study showed that the tensile stress components develop cusp-like structures for $\Wi>0.5$; they grow unbounded in time for $\Wi>0.9$. This behavior is attributed to the known flaw of the Oldroyd-B model to assume infinitely extensional Hookean springs. For lower Weissenberg numbers, the stress field is bounded and sufficiently regular for a convergence analysis \cite{Pimenta2017}.

\begin{figure}[H]
		\subfigure{
					\begin{tikzpicture}
					\begin{loglogaxis}
					[
					xlabel=\small{$h/\br{2\pi}$},
					ylabel=\small{$\abs{\chi_{11,0}\br{h} - \chi_{11,0}\br{\te{extr.}}}$},
					width=0.45\textwidth,
				    height=0.45\textwidth,
					grid style=off,
					ticklabel style = {font=\small},
					legend style={at={(0,1)},anchor=north west,font=\small},
					]
					\addplot[color=blue, mark=*] table [x=h, y=err_t110, col sep=comma] {fig/PeriodicBox/EOC_PeriodicBox.formGLS2.lambda0.1000.csv};
					\addlegendentry{SUPG, $n=2.012$};								
					\addplot[color=orange, mark=*] table [x=h, y=err_t110, col sep=comma] {fig/PeriodicBox/EOC_PeriodicBox.formGLS3.lambda0.1000.csv};
					\addlegendentry{GLS, $n=1.99$};					
					\addplot[color=magenta, mark=*] table [x=h, y=err_t110, col sep=comma] {fig/PeriodicBox/EOC_PeriodicBox.formASS.lambda0.1000.csv};
					\addlegendentry{ASGS, $n=1.975$};					
					\end{loglogaxis}
					\end{tikzpicture}
		}~
		\subfigure{			
					\begin{tikzpicture}
					\begin{loglogaxis}
					[
					xlabel=\small{$h/\br{2\pi}$},
					ylabel=\small{$\abs{\varepsilon_0\br{h} - \varepsilon_0\br{\te{extr.}}}$},
					grid style=off,
					width=0.45\textwidth,
				    height=0.45\textwidth,
					ticklabel style = {font=\small},
					legend style={at={(0.1,1)},anchor=north west,font=\small},
					]
					\addplot[color=blue, mark=*] table [x=h, y=err_u1x10, col sep=comma] {fig/PeriodicBox/EOC_PeriodicBox.formGLS2.lambda0.1000.csv};
					\addlegendentry{SUPG, $n=2.004$};								
					\addplot[color=orange, mark=*] table [x=h, y=err_u1x10, col sep=comma] {fig/PeriodicBox/EOC_PeriodicBox.formGLS3.lambda0.1000.csv};
					\addlegendentry{GLS, $n=1.973$};					
					\addplot[color=magenta, mark=*] table [x=h, y=err_u1x10, col sep=comma] {fig/PeriodicBox/EOC_PeriodicBox.formASS.lambda0.1000.csv};
					\addlegendentry{ASGS, $n=2.001$};					
					\end{loglogaxis}
					\end{tikzpicture}
		}
\caption{Convergence rates of first normal stress component $\chi_{11}$, and rate of strain $\varepsilon_0$ at $\Wi=0.15$.}
\label{fig:pboxconvrates}
\end{figure}
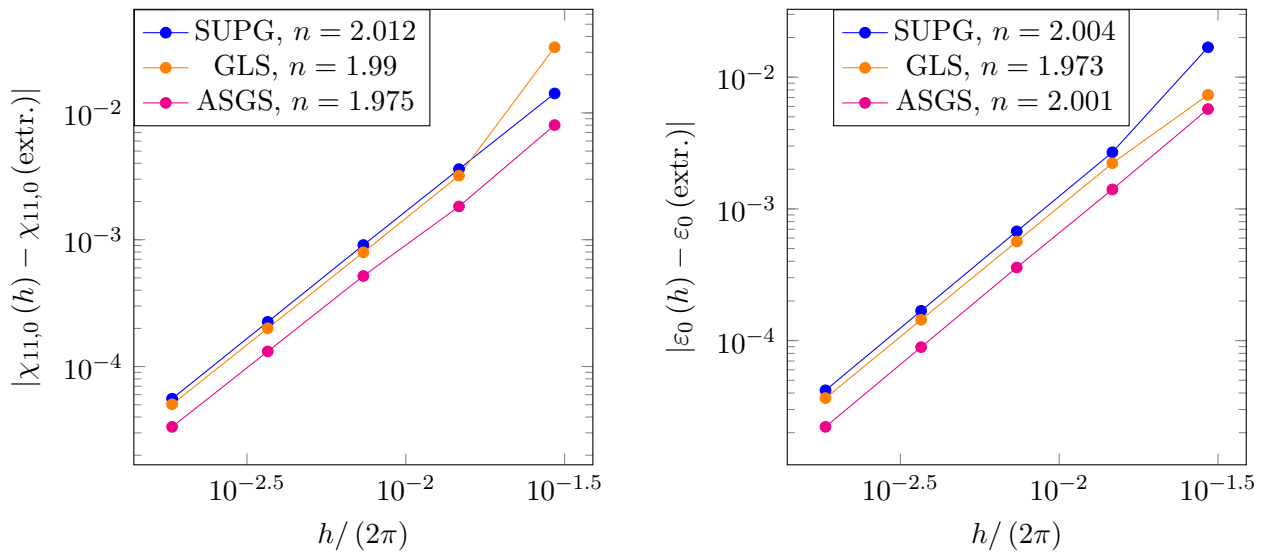

We take $\Wi=0.15$ in our study and use quadrilateral elements of side length $h$. To obtain estimates for convergence rates, we use Richardson extrapolation \cite{Richardson1911}. We analyze the strain rate $\varepsilon_0=\partial u / \partial x |_{\br{0,0}}$, and the first normal stress component $\chi_{11,0}$ at the center of the box, where the highest generated stresses are located.
We use five subsequently refined meshes, where for each refinement level, the element size is cut in half. The coarsest mesh consists of $34\times34$ elements, i.e., $h\approx0.1848$. Based on the three finest meshes, the experimental order of convergence $n$, e.g., for $\chi_{11,0}$, can be computed as
\begin{align}
n=\log{\br{\frac{\chi_{11,0}\br{2h} - \chi_{11,0}\br{h}}{\chi_{11,0}\br{4h} - \chi_{11,0}\br{2h}}}}/\log(0.5)\,,
\end{align}
and the extrapolated value as
\begin{align}
\chi_{11,0}\br{\te{extr.}} = \frac{2^n \chi_{11,0}\br{h} - \chi_{11,0}\br{2h}}{2^n-1}\,.
\end{align}
Fig.~\ref{fig:pboxconvrates} displays the absolute errors w.r.t to the extrapolated values. For all methods, we observe an approximate second-order convergence in both quantities. 

\subsection{Flow past a cylinder}
\label{sec:cylinder}

We present the results for the two-dimensional flow past a cylinder in a channel with a ratio of two between channel height and cylinder radius. The agreement of the numerical data between independent studies makes this test case a most frequent choice for assessment of accuracy and robustness of numerical methods for viscoelastic flows \cite{Alves2001,Hulsen2005,Knechtges2014a,Coronado2007,Fan99}. Our results are compared to available data in the literature.

\begin{figure}[H]
\centering
\def\svgwidth{1.0\textwidth}
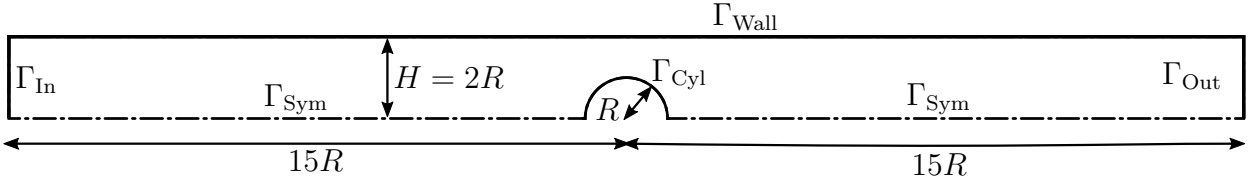
\caption{Schematic drawing of the cylinder test case with $H/R=2$.}
\label{fig:schematicCyl2to1}
\end{figure}

We consider half of the domain in our computation as the problem is symmetric. Fig.~(\ref{fig:schematicCyl2to1}) shows the description of the problem schematically. A no-slip boundary condition is prescribed at the wall $\Gamma_{\te{Sym}}$ and on the cylinder $\Gamma_{\te{Cyl}}$, i.e., $\vect{u}=\vect{0}$. Symmetry conditions on $\Gamma_{\te{Sym}}$ require us to take $\vect{u} \mydot \vect{n} = 0$ , and to take the tangential component of the traction $\ten{T}\mydot \vect{n}$ as zero, thus leading to $u_2=0$ and $\chi_{12}=0$. At the outlet $\Gamma_{\te{Out}}$, the vertical component of the velocity is taken as zero. To avoid large computational domains, we have imposed the fully developed stress and velocity profiles for an Oldroyd-B flow on the inlet $\Gamma_{\te{In}}$:
\begin{align}
u_1 &= \frac{3}{2} \overline{u} \br{1 - \br{\frac{y}{H}}^2} \, ,\, u_2 = 0\\
\sigma_{11} &= 18 \lambda (1-\beta) \etaT \overline{u}^2 \frac{y^2}{H^4} \, ,\, \sigma_{12} = -3 (1-\beta) \etaT \overline{u} \frac{y}{H^2} \, ,\,\sigma_{22} = 0 \,.
\end{align}
Alternatively, imposing zero stresses is also possible. The extra stresses $\ten{\sigma}$ are transformed to $\ten{\chi}$ according to the inverted relation of (\ref{eqn:transformsigchi}) for the inlet values, $\ten{\chi}=\frac{\etaP}{\lambda}\log{\br{\frac{\lambda}{\etaP}\ten{\sigma}+\ten{I}}}$. The coordinate system is located at the center of the cylinder. The parameters for the test case are $\etaT=1$, $\beta=0.59$, $\overline{u}=2$, $R=1$, and $H=2$. A creeping flow is realized in our implementation by taking $\rho=0$. The relaxation is chosen to obtain a Weissenberg number defined by $\te{Wi}=\lambda \overline{u} / R$.\\

\begin{figure}[H]
\includegraphics[trim=0 0 0 0, clip, width=1.0\textwidth]{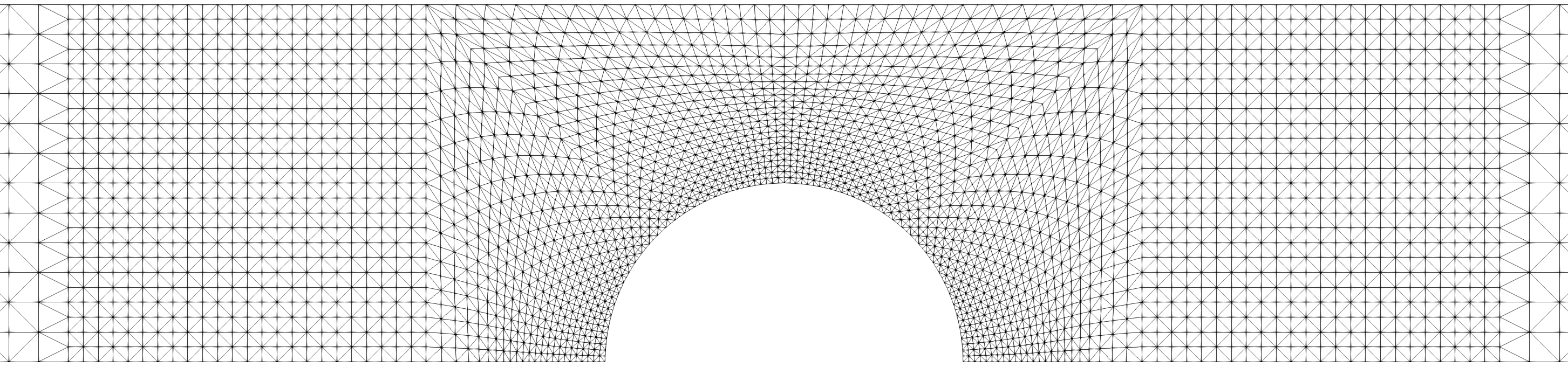}
\caption{Triangular mesh T1 for the flow past a cylinder}
\label{fig:cylmesh}
\end{figure}

Tab.~\ref{tab:cylmesh} summarizes the main characteristics of the computational meshes. We employed triangular meshes with alternating structure with a decreasing element size towards the cylinder (see Fig.~\ref{fig:cylmesh}).

\begin{table}[H]
\centering
\caption{Mesh characteristics for the flow past a cylinder}
\label{tab:cylmesh}
\begin{tabular}{c c c c}
\hline
Mesh & Number of elements & Number of nodes & Smallest element on cylinder \\
\hline
T1 & 10104 & 5341 & 0.0327 \\
T2 & 40368 & 20761 & 0.0164 \\
T3 & 161376 & 81841 & 0.0082 \\
T4 & 645312 & 324961 & 0.0041 \\
\hline
\end{tabular}
\end{table}

\subsubsection{Convergence assessment of Newton method: semi-analytic approximation vs. finite differences for computation of Jacobians}
\label{sec:anavsfd}

In optimal conditions, the Newton-Raphson method converges locally at a rate of two, thus, in terms of the residual, the inequality
\begin{align}
\norm{\ten{Res}\br{U_h^{i+1}}} \leq M \norm{\ten{Res}\br{U_h^{i}}}^2
\end{align}
holds for a positive constant $M$.

We compare two different implementations of the Jacobians $\frac{\partial \ten{Res}}{\partial U_n}$ for the SUPG method (\ref{eqn:glssupg}),(\ref{eqn:projsupg}). Besides the numerical approximation by using central finite differences (\ref{eqn:centralfdscheme}), we implemented analytic Jacobians, where the stabilization parameters, as well as the advective velocity, are computed with the solution of the previous iteration $U_h^{i-1}$. This approach will be referred to as semi-analytic. The threshold parameter to decide convergence of the non-linear iteration is set to $\epsilon=10^{-10}$.

\begin{figure}[H]
					\begin{tikzpicture}
					\begin{loglogaxis}
					[
					xlabel=\small{$\norm{\ten{Res}\br{U_h^{i}}} / \sqrt{\ndof}$},
					ylabel=\small{$\norm{\ten{Res}\br{U_h^{i+1}}} / \sqrt{\ndof}$},
					grid style=off,
					ticklabel style = {font=\small},
					legend style={at={(1.1,1)},anchor=north west,font=\small},
					]
					\addplot[color = blue, mark=x, thick, dashed] table [x=Res0, y=Res1, col sep=comma] {fig/Cylinder2to1/LogLogResidual_GLS_Ref3_lambda0.1500.csv};\addlegendentry{semi-analytic, $\Wi=0.3$};			
					\addplot[color = green, mark=x, thick, dashed] table [x=Res0, y=Res1, col sep=comma] {fig/Cylinder2to1/LogLogResidual_GLS_Ref3_lambda0.3000.csv};\addlegendentry{semi-analytic, $\Wi=0.6$};	
					\addplot[color = orange, mark=x, thick, dashed] table [x=Res0, y=Res1, col sep=comma] {fig/Cylinder2to1/LogLogResidual_GLS_Ref3_lambda0.4500.csv};\addlegendentry{semi-analytic, $\Wi=0.9$};	
					
					\addplot[color = blue, mark=o, thick, solid] table [x=Res0, y=Res1, col sep=comma] {fig/Cylinder2to1/LogLogResidual_GLS2_Ref3_lambda0.1500.csv};\addlegendentry{central f.d., $\Wi=0.3$};			
					\addplot[color = green, mark=o, thick, solid] table [x=Res0, y=Res1, col sep=comma] {fig/Cylinder2to1/LogLogResidual_GLS2_Ref3_lambda0.3000.csv};\addlegendentry{central f.d., $\Wi=0.6$};	
					\addplot[color = orange, mark=o, thick, solid] table [x=Res0, y=Res1, col sep=comma] {fig/Cylinder2to1/LogLogResidual_GLS2_Ref3_lambda0.4500.csv};\addlegendentry{central f.d., $\Wi=0.9$};	

					\addplot[color=black, solid, mark=none]
					coordinates{ (1e-7,1e-12) (1e-2,1e-2)}; \addlegendentry{slope 2};
					
					\addplot[color=black, solid, mark=none]
					coordinates{ (1e-10,1e-11) (1e-2,1e-3)}; \addlegendentry{slope 1};

					\end{loglogaxis}
					\end{tikzpicture}
\caption{Residual of $\br{i+1}$-th iteration vs. residual of $i$-th iteration for mesh T2 for different Weissenberg numbers. Comparison of semi-analytic Jacobian vs. finite difference Jacobian.}
\label{fig:fdWi}
\end{figure}
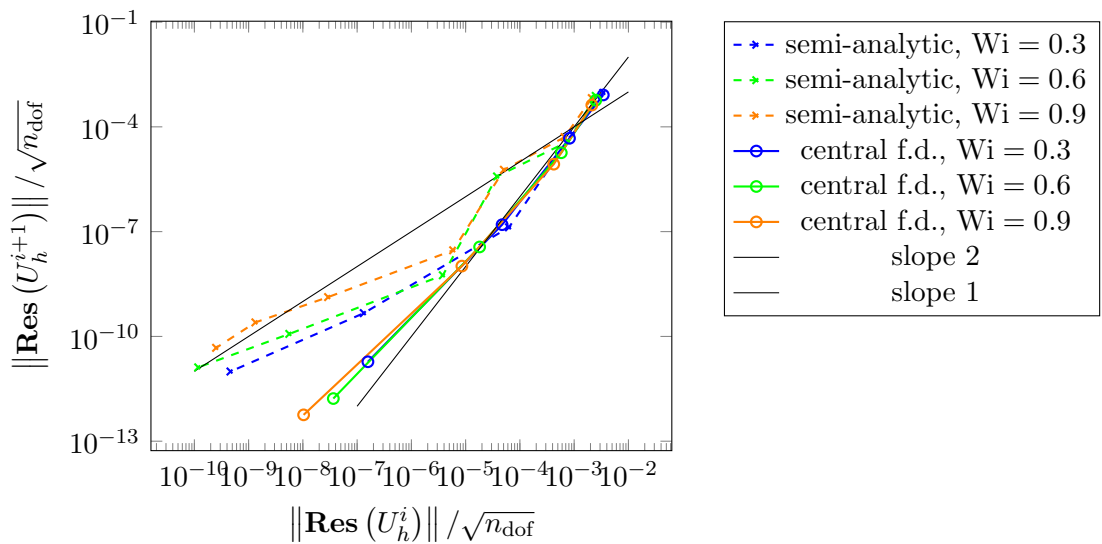

To visualize the local rate of convergence, we show the residual of the $\br{i+1}$-th versus the $i$-th iteration in a logarithmic scaling for different Weissenberg numbers (Fig.~\ref{fig:fdWi}). The graphs are constructed by the residuals in descending order of the iterations; the first data point in the figure is connected to the last iteration step. The finite difference approach closely follows a second-order rate in the earlier iterations. Merely in the last iteration, the slope becomes suboptimal, after comparing to the reference slope. This slight deterioration might be attributed to higher errors in the solution of the linear problem \cite{Knechtges2015}. The semi-analytic approach already clearly deviates from a second order after the first iterations. Nevertheless, the local rates still exceed one.

Tab.~\ref{tab:cyllimwi} collects the critical Weissenberg numbers for all methods of all meshes at which a breakdown occurred, either due to a blow-up of the residual, not meeting the convergence criterion, or not obtain a converged solution of the linear iterative solver. The values are accurate up to an uncertainty of the last incremental step in the Weissenberg number $\Delta \lambda=0.01$.
Across all meshes, the critical number was higher for the finite difference approach than the semi-analytic approach.
Therefore, we believe the reduced convergence rate might also be connected to the robustness.

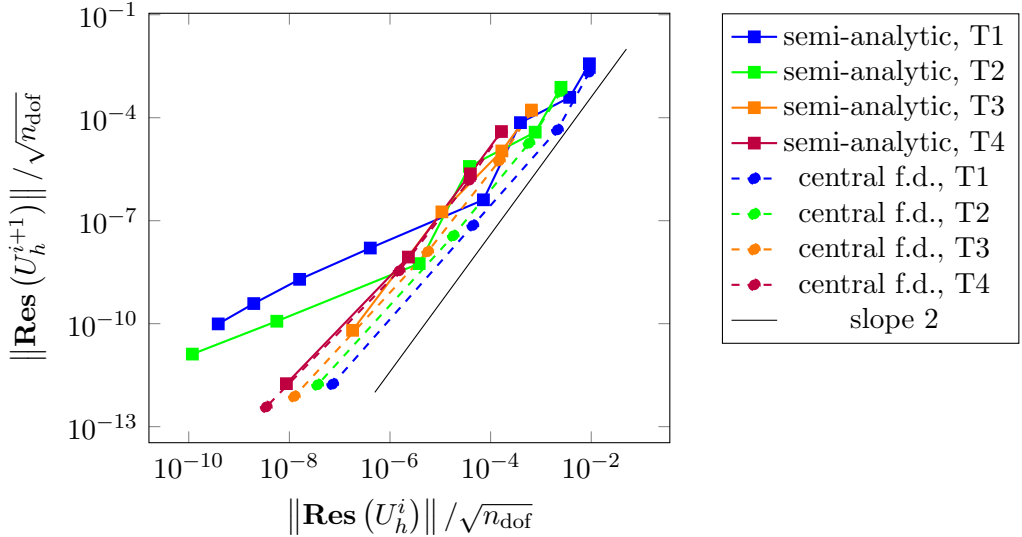
\begin{figure}[H]
					\begin{tikzpicture}
					\begin{loglogaxis}
					[
					xlabel=\small{$\norm{\ten{Res}\br{U_h^{i}}} / \sqrt{\ndof}$},
					ylabel=\small{$\norm{\ten{Res}\br{U_h^{i+1}}} / \sqrt{\ndof}$},
					grid style=off,
					ticklabel style = {font=\small},
					legend style={at={(1.1,1)},anchor=north west,font=\small},
					]
					\addplot[color = blue, mark=square*, thick, solid] table [x=Res0, y=Res1, col sep=comma] {fig/Cylinder2to1/LogLogResidual_GLS_Ref2_lambda0.3000.csv};\addlegendentry{semi-analytic, T1};			
					\addplot[color = green, mark=square*, thick, solid] table [x=Res0, y=Res1, col sep=comma] {fig/Cylinder2to1/LogLogResidual_GLS_Ref3_lambda0.3000.csv};\addlegendentry{semi-analytic, T2};	
					\addplot[color = orange, mark=square*, thick, solid] table [x=Res0, y=Res1, col sep=comma] {fig/Cylinder2to1/LogLogResidual_GLS_Ref4_lambda0.3000.csv};\addlegendentry{semi-analytic, T3};	
					\addplot[color = purple, mark=square*, thick, solid] table [x=Res0, y=Res1, col sep=comma] {fig/Cylinder2to1/LogLogResidual_GLS_Ref5_lambda0.3000.csv};\addlegendentry{semi-analytic, T4};	

					\addplot[color = blue, mark=*, thick, dashed] table [x=Res0, y=Res1, col sep=comma] {fig/Cylinder2to1/LogLogResidual_GLS2_Ref2_lambda0.3000.csv};\addlegendentry{central f.d., T1};			
					\addplot[color = green, mark=*, thick, dashed] table [x=Res0, y=Res1, col sep=comma] {fig/Cylinder2to1/LogLogResidual_GLS2_Ref3_lambda0.3000.csv};\addlegendentry{central f.d., T2};	
					\addplot[color = orange, mark=*, thick, dashed] table [x=Res0, y=Res1, col sep=comma] {fig/Cylinder2to1/LogLogResidual_GLS2_Ref4_lambda0.3000.csv};\addlegendentry{central f.d., T3};	
					\addplot[color = purple, mark=*, thick, dashed] table [x=Res0, y=Res1, col sep=comma] {fig/Cylinder2to1/LogLogResidual_GLS2_Ref5_lambda0.3000.csv};\addlegendentry{central f.d., T4};	
					
					\addplot[color=black, solid, mark=none]
					coordinates{ (0.5e-6,1e-12) (0.5e-1,1e-2)}; \addlegendentry{slope 2};
					
					
					\end{loglogaxis}
					\end{tikzpicture}
\caption{Residual of $\br{i+1}$-th iteration vs. residual of $i$-th iteration for a Weissenberg number $\Wi=0.6$ on different meshes. Comparison of semi-analytic Jacobian vs. finite difference Jacobian.}
\label{fig:fdmeshes}
\end{figure}

Fig.~\ref{fig:fdmeshes} shows a comparative study between all meshes for a single Weissenberg number. 	The semi-analytic approach reveals a mesh-dependent behavior. Whereas the calculations on coarser grids deviate more clearly from the values of the finite difference calculations, the residuals of the finer grids for both approaches lie almost on top of each other. 

This behavior suggests two conclusions: (1) As the stabilization parameter depends on the mesh size, we hypothesize that the simplification in the stabilization parameter of the semi-analytic approach causes the deterioration of convergence we observed. The finite difference approach does not seem to be significantly affected by the mesh resolution. (2) The residuals on the finer meshes drop similarly fast, i.e., the finite difference approach behaves comparably to the semi-analytic approach, making the finite difference approach an accurate and easily adaptable alternative.

The study seems to support the thesis that simplifications in the non-linearities of the stabilization terms may impair the convergence of Newton's method.

\subsubsection{Step size dependence}

\begin{figure}[H]
					\begin{tikzpicture}
					\begin{semilogyaxis}
					[
					xlabel=\small{$\te{number of iterations}$},
					ylabel=\small{$\norm{R}_2/\sqrt{N_{\te{dof}}}$},
					grid style=off,
					ticklabel style = {font=\small},
					legend style={at={(1.1,1)},anchor=north west,font=\small},
					cycle list name=color
					]
					\addplot[color=black, mark=square*, thick, solid] table [x=iit, y=Res, col sep=comma] {fig/Cylinder2to1/Residual_GLS2_Ref3_lambda0.1500_fdeps2.csv};\addlegendentry{$\Delta_{\te{FD}}=10^{-2}$};			
					\addplot[color=blue, mark=square*, thick, solid] table [x=iit, y=Res, col sep=comma] {fig/Cylinder2to1/Residual_GLS2_Ref3_lambda0.1500_fdeps3.csv};\addlegendentry{$\Delta_{\te{FD}}=10^{-3}$};			
					\addplot[color=gray, mark=square*, thick, solid] table [x=iit, y=Res, col sep=comma] {fig/Cylinder2to1/Residual_GLS2_Ref3_lambda0.1500_fdeps4.csv};\addlegendentry{$\Delta_{\te{FD}}=10^{-4}$};			
					\addplot[color=orange, mark=square*, thick, solid] table [x=iit, y=Res, col sep=comma] {fig/Cylinder2to1/Residual_GLS2_Ref3_lambda0.1500_fdeps5.csv};\addlegendentry{$\Delta_{\te{FD}}=10^{-5}$};			
					\addplot[color=red, mark=square*, thick, solid] table [x=iit, y=Res, col sep=comma] {fig/Cylinder2to1/Residual_GLS2_Ref3_lambda0.1500_fdeps6.csv};\addlegendentry{$\Delta_{\te{FD}}=10^{-6}$};			
					\addplot[color=magenta, mark=square*, thick, solid] table [x=iit, y=Res, col sep=comma] {fig/Cylinder2to1/Residual_GLS2_Ref3_lambda0.1500_fdeps7.csv};\addlegendentry{$\Delta_{\te{FD}}=10^{-7}$};			
					\addplot[color=purple, mark=square*, thick, solid] table [x=iit, y=Res, col sep=comma] {fig/Cylinder2to1/Residual_GLS2_Ref3_lambda0.1500_fdeps8.csv};\addlegendentry{$\Delta_{\te{FD}}=10^{-8}$};		
					\addplot[color=brown, mark=square*, thick, solid] table [x=iit, y=Res, col sep=comma] {fig/Cylinder2to1/Residual_GLS2_Ref3_lambda0.1500_fdeps9.csv};\addlegendentry{$\Delta_{\te{FD}}=10^{-9}$};		
					\addplot[color=green, mark=square*, thick, solid] table [x=iit, y=Res, col sep=comma] {fig/Cylinder2to1/Residual_GLS2_Ref3_lambda0.1500_fdeps10.csv};\addlegendentry{$\Delta_{\te{FD}}=10^{-10}$};	
					\addplot[color=yellow, mark=square*, thick, solid] table [x=iit, y=Res, col sep=comma] {fig/Cylinder2to1/Residual_GLS2_Ref3_lambda0.1500_fdeps11.csv};\addlegendentry{$\Delta_{\te{FD}}=10^{-11}$};	
					\addplot[color=olive, mark=square*, thick, solid] table [x=iit, y=Res, col sep=comma] {fig/Cylinder2to1/Residual_GLS2_Ref3_lambda0.1500_fdeps12.csv};\addlegendentry{$\Delta_{\te{FD}}=10^{-12}$};	
					\end{semilogyaxis}
					\end{tikzpicture}
\caption{Residual vs. number of iterations of the non-linear iteration for different step sizes of the finite difference scheme at $\Wi=0.3$ for mesh T2.}
\label{fig:fdstepsize}
\end{figure}
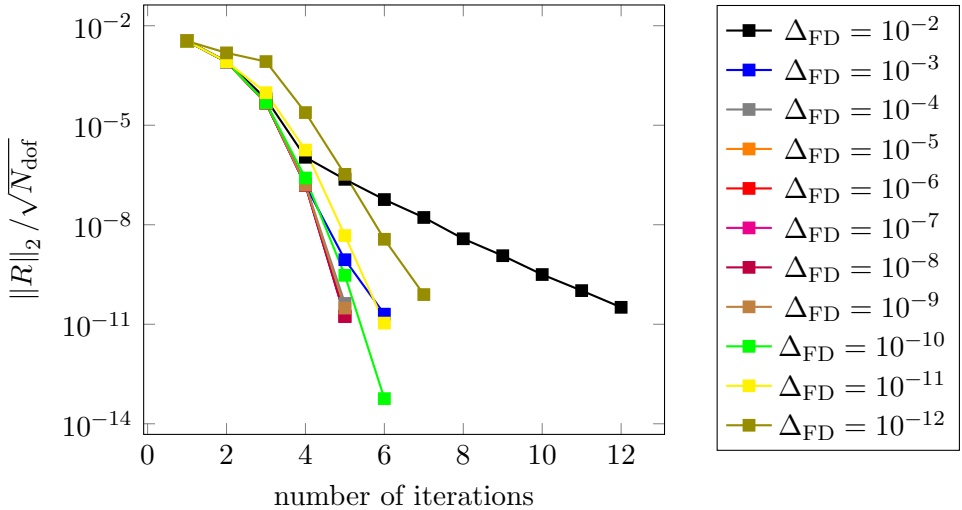

Two aspects need to be taken into account when choosing a step size for a finite-difference stencil: (1) A step size too small leads to a loss of significant digits as two almost equal numbers are subtracted; (2) A step size too large introduces a higher truncation error of the stencil.
A wrong choice of the step size may result in sub-optimal convergence or divergence of the Newton method. For the central finite difference stencil, the minimum error is obtained by taking $\Delta_{\te{FD}}\approx10^{-5}$, which is about the cubic root of the machine epsilon in double precision \cite{Sauer2018}, p.248.
In this section, we study how the step size $\Delta_{\te{FD}}$ affects the convergence behavior of the non-linear iteration.
Fig.~\ref{fig:fdstepsize} depicts the behavior of the residual over the number of iterations of the SUPG method on the mesh T2.
For $\Delta_{\te{FD}}\geq10^{-3}$ and $\Delta_{\te{FD}}\leq10^{-9}$, the residual drops at a lower rate than for the remaining values. For values $\Delta_{\te{FD}}\leq10^{-13}$, the linear solver diverged. We think both limitations are attributed to either the truncation error of the stencil or a loss of significant digits. We could not observe significant differences in the computed results, convergence behavior, or limiting Weissenberg number by taking the step size within the range $10^{-8}\leq\Delta_{\te{FD}}\leq10^{-4}$. Further numerical experiments on different meshes substantiated our findings.

\subsubsection{Drag predictions and limiting Weissenberg numbers}

The breakdown of a computation for the flow past a cylinder was limited to Weissenberg numbers between $0.7$ and $1.0$ for standard formulations \cite{Coronado2005a}, \cite{Fan99}, \cite{Alves2001}. 
The increased stability of logarithmic conformation formulations has been shown in several folllowing works \cite{Hulsen2005}, \cite{Coronado2007}, \cite{Knechtges2014a}, \cite{Moreno2019}, \cite{Afonso2009}, \cite{Damanik2010}, which allowed computations beyond $\Wi=1.0$. 
Tab.~\ref{tab:dragliterature} contrasts our drag coefficient values for all methods with available literature data.
The non-dimensional drag coefficient value is defined by
\begin{align}
K = - \frac{2}{\etaT \overline{u}} \int_{\Gamma_{\te{Cyl}}} \vect{e}_x \mydot \ten{T} \mydot \vect{n} \,.
\end{align}
Across all presented methods, we could not observe considerable differences. The extrapolated (extr.) values for the ASGS method, obtained by Richardson extrapolations, are shown for all cases where we obtained a positive extrapolated convergence order. Whereas the extrapolated values coincide with five digits to the reference data up to $\Wi=0.9$, the values on the most refined mesh are less exact. This behavior suggests our meshes were not sufficiently refined to yield fully converged results. Beyond that value, our methods seem to underestimate other reported predictions. For Weissenberg numbers in this range, the values among the published works are no longer matching.

Although we could compute steady-state solutions beyond $\Wi=0.9$, it cannot be concluded that physical stationary solutions exist.
The determination of the transition Weissenberg number from a steady-state to a transient solution remains an open debate \cite{Alves2021}.
Afonso \cite{Afonso2009} reported not being able to obtain steady-state solutions above $0.9$. Fan et al. \cite{Fan99} conjectured results beyond $0.8$ might be numerical artifacts.

Despite the questionable existence of steady-state solutions, we may draw attention to Tab.~\ref{tab:cyllimwi} again. Regardless of the method, the log-conf reformulation alleviates the HWNP. The limiting Weissenberg numbers for the finer meshes are well beyond one, and a severe mesh dependence of the limiting Weissenberg number does not seem to be as evident as opposed to standard formulations. Whereas the GLS and SUPG limiting numbers are comparable, the ASGS method displays, by far, the highest limiting Weissenberg numbers.

\begin{table}[H]
\caption{Limiting Weissenberg numbers for the flow past a cylinder of presented methods and meshes.}
\label{tab:cyllimwi}
\begin{tabular}{l|llll}
                     & T1   & T2   & T3   & T4   \\
\hline
SUPG (semi-anal. jac.) & 1.04 & 1.00 & 1.60 & 1.40 \\
SUPG                 & 1.09 & 1.52 & 1.61 & 2.00 \\
GLS                  & 1.09 & 1.51 & 1.65 & 1.96 \\
ASGS                 & 1.19 & 3.01 & 2.89 & 2.72
\end{tabular}
\end{table}
				
\begin{table}[H]
\caption{Drag coefficients of the cylinder with mesh T4 compared to literature data.}
\label{tab:dragliterature}
\begin{tabular}{lllllllll}
\hline
Wi  & SUPG     & GLS      & ASGS   & ASGS (extr.)  & \cite{Hulsen2005} & \cite{Fan99} (MIX1) & \cite{Afonso2009} & \cite{Moreno2019} \\
\hline
0.0 & 132.298 & 132.293 & 132.287 &   132.359      & 132.358          &        132.36           &                                     &                                     \\
0.1 & 130.329 & 130.335 & 130.322 &  130.367    & 130.363             &        130.36        &                                     & 130.30                              \\
0.2 & 126.617 & 126.629 & 126.619 &                   & 126.6226             &       126.62        &                                     &                                     \\
0.3 & 123.202 & 123.217 & 123.211 &  123.196     & 123.193             &        123.19         &                                     &                                     \\
0.4 & 120.617 & 120.634 & 120.632 &  120.5995   & 120.596              &        120.59       &                                     &                                     \\
0.5 & 118.865 & 118.882 & 118.883 &  118.836     & 118.836               &        118.83      & 118.781                             & 118.82                              \\
0.6 & 117.826 & 117.842 & 117.847 &  117.785     & 117.792                &      117.77       & 117.778                             &                                     \\
0.7 & 117.377 & 117.392 & 117.399 &  117.325     & 117.34           &        117.32           & 117.350                             &                                     \\
0.8 & 117.413 & 117.427 & 117.436 &  117.3595   & 117.373           &        117.36          & 117.380                             &                                     \\
0.9 & 117.832 & 117.845 & 117.854 &  117.793   & 117.787             &       117.79         & 117.797                             &                                     \\
1.0 & 118.530 & 118.539 & 118.552 &  118.512 & 118.501         &       118.49             & 118.662                             & 118.88                              \\
1.1 & 119.464 & 119.462 & 119.481 &  119.461     & 119.466           &                  & 119.740                             &                                     \\
1.2 & 120.609 & 120.592 & 120.617 &  120.613     & 120.65               &               & 120.985                             & 121.14                              \\
1.3 & 121.943 & 121.911 & 121.941 &                &                           &                  &                                     &                                     \\
1.4 & 123.446 & 123.404 & 123.441 &                & 123.587             &                & 124.129                             & 124.14                              \\
1.5 & 125.106 & 125.057 & 125.098 &                &                            &         & 126.022                             &                                     \\
1.6 & 126.910 & 126.861 & 126.898 &                & 127.172              &               & 127.759                             & 127.66                              \\
1.7 & 128.847 & 128.804 & 128.825 &                &                           &          & 130.012                             &                                     \\
1.8 & 130.909 & 130.881 & 130.874 &                & 131.285            &                 & 132.024                             & 131.53                              \\
1.9 & 133.084 & 133.088 & 133.041 &                &                          &           & 134.188                             &                                     \\
2.0 & 135.372 &               & 135.320 &                &                          &           & 136.580                             & 135.53                              \\
2.1 &               &               & 137.704 &                &                           &          &                                     &                                     \\
2.2 &               &               & 140.183 &                &                          &           & 141.801                             & 139.62                              \\
2.3 &               &               & 142.749 &                &                         &            &                                     &                                     \\
2.4 &               &               & 145.390 &                &                         &            & 146.730                             & 143.66                              \\
2.5 &               &               & 148.096 &                &                         &            & 149.112                             &                                    
\end{tabular}
\end{table}

\subsubsection{Convergence of stresses}

\begin{figure}[H]				
					\subfigure{
					\begin{tikzpicture}
					\begin{axis}
					[
					set layers,					
					mark layer=axis background,
					width=0.45\textwidth,
				    height=0.45\textwidth,
					xlabel=\small{$s$},
					ylabel=\small{$\sigma_{11}/ \br{\etaT \, \overline{u} \, / \, R}$},
					grid style=off,
					ticklabel style = {font=\small},
					legend style={at={(1.0,1.0)},anchor=north east,font=\small},
					xtick={0,1,2,3,4,5,6},
					]		
					
					\addplot[color = black, mark=none*, style=thick, solid] table [x=s, y=txx, col sep=comma] {fig/Cylinder2to1_Lineplots/Cylinder2to1_Lineplot_ASS_T2_Wi0.6000.csv};\addlegendentry{ASGS, T1};		
					
					\addplot[color = blue, mark=none*, style=thick, solid] table [x=s, y=txx, col sep=comma] {fig/Cylinder2to1_Lineplots/Cylinder2to1_Lineplot_ASS_T3_Wi0.6000.csv};\addlegendentry{ASGS, T2};		
					
					\addplot[color = green, mark=none*, style=thick, solid] table [x=s, y=txx, col sep=comma] {fig/Cylinder2to1_Lineplots/Cylinder2to1_Lineplot_ASS_T4_Wi0.6000.csv};\addlegendentry{ASGS, T3};		
					
					\addplot[color = purple, mark=none*, style=thick, solid] table [x=s, y=txx, col sep=comma] {fig/Cylinder2to1_Lineplots/Cylinder2to1_Lineplot_ASS_T5_Wi0.6000.csv};\addlegendentry{ASGS, T4};		
					
					\addplot[color = black, mark=none*, style=thick, dashed] table [x=s, y=txx, col sep=comma] {fig/Cylinder2to1_Lineplots/Cylinder2to1_Lineplot_GLS3_T2_Wi0.6000.csv};\addlegendentry{GLS, T1};		
					
					\addplot[color = blue, mark=none*, style=thick, dashed] table [x=s, y=txx, col sep=comma] {fig/Cylinder2to1_Lineplots/Cylinder2to1_Lineplot_GLS3_T3_Wi0.6000.csv};\addlegendentry{GLS, T2};		
					
					\addplot[color = green, mark=none*, style=thick, dashed] table [x=s, y=txx, col sep=comma] {fig/Cylinder2to1_Lineplots/Cylinder2to1_Lineplot_GLS3_T4_Wi0.6000.csv};\addlegendentry{GLS, T3};		
					
					\addplot[color = purple, mark=none*, style=thick, dashed] table [x=s, y=txx, col sep=comma] {fig/Cylinder2to1_Lineplots/Cylinder2to1_Lineplot_GLS3_T5_Wi0.6000.csv};\addlegendentry{GLS, T4};			
					
					\addplot[color = orange, mark=pentagon, mark options={scale=0.9}, only marks, thick] table [x=s, y=txx, col sep=comma] {fig/Cylinder2to1_Lineplots/References/Cylinder2to1_Lineplot_Damanik2010_Wi0.6000.csv};\addlegendentry{\cite{Damanik2010}};				
					
					\end{axis}
					\end{tikzpicture}	
					}~
					\subfigure{
					\begin{tikzpicture}
					\begin{axis}
					[
					set layers,					
					mark layer=axis background,
					width=0.45\textwidth,
				    height=0.45\textwidth,
					xlabel=\small{$s$},
					ylabel=\small{$\sigma_{11}/ \br{\etaT \, \overline{u} \, / \, R}$},
					grid style=off,
					ticklabel style = {font=\small},
					xtick={0,1,2,3,4,5,6},
					xmin=3.1415926536,
					xmax=6,
					]		
					
					\addplot[color = black, mark=none*, style=thick, solid] table [x=s, y=txx, col sep=comma] {fig/Cylinder2to1_Lineplots/Cylinder2to1_Lineplot_ASS_T2_Wi0.6000.csv};
					
					\addplot[color = blue, mark=none*, style=thick, solid] table [x=s, y=txx, col sep=comma] {fig/Cylinder2to1_Lineplots/Cylinder2to1_Lineplot_ASS_T3_Wi0.6000.csv};
					
					\addplot[color = green, mark=none*, style=thick, solid] table [x=s, y=txx, col sep=comma] {fig/Cylinder2to1_Lineplots/Cylinder2to1_Lineplot_ASS_T4_Wi0.6000.csv};
					
					\addplot[color = purple, mark=none*, style=thick, solid] table [x=s, y=txx, col sep=comma] {fig/Cylinder2to1_Lineplots/Cylinder2to1_Lineplot_ASS_T5_Wi0.6000.csv};
					
					\addplot[color = black, mark=none*, style=thick, dashed] table [x=s, y=txx, col sep=comma] {fig/Cylinder2to1_Lineplots/Cylinder2to1_Lineplot_GLS3_T2_Wi0.6000.csv};
					
					\addplot[color = blue, mark=none*, style=thick, dashed] table [x=s, y=txx, col sep=comma] {fig/Cylinder2to1_Lineplots/Cylinder2to1_Lineplot_GLS3_T3_Wi0.6000.csv};
					
					\addplot[color = green, mark=none*, style=thick, dashed] table [x=s, y=txx, col sep=comma] {fig/Cylinder2to1_Lineplots/Cylinder2to1_Lineplot_GLS3_T4_Wi0.6000.csv};
					
					\addplot[color = purple, mark=none*, style=thick, dashed] table [x=s, y=txx, col sep=comma] {fig/Cylinder2to1_Lineplots/Cylinder2to1_Lineplot_GLS3_T5_Wi0.6000.csv};
					
					\addplot[color = orange, mark=pentagon, mark options={scale=0.9}, only marks, thick] table [x=s, y=txx, col sep=comma] {fig/Cylinder2to1_Lineplots/References/Cylinder2to1_Lineplot_Damanik2010_Wi0.6000.csv};
					
					\end{axis}
					\end{tikzpicture}	
					}					
					
\caption{Stress profiles on the cylinder wall and wake at $\te{Wi}=0.6$}
\label{fig:cyllineplotmeshes06}
\end{figure}
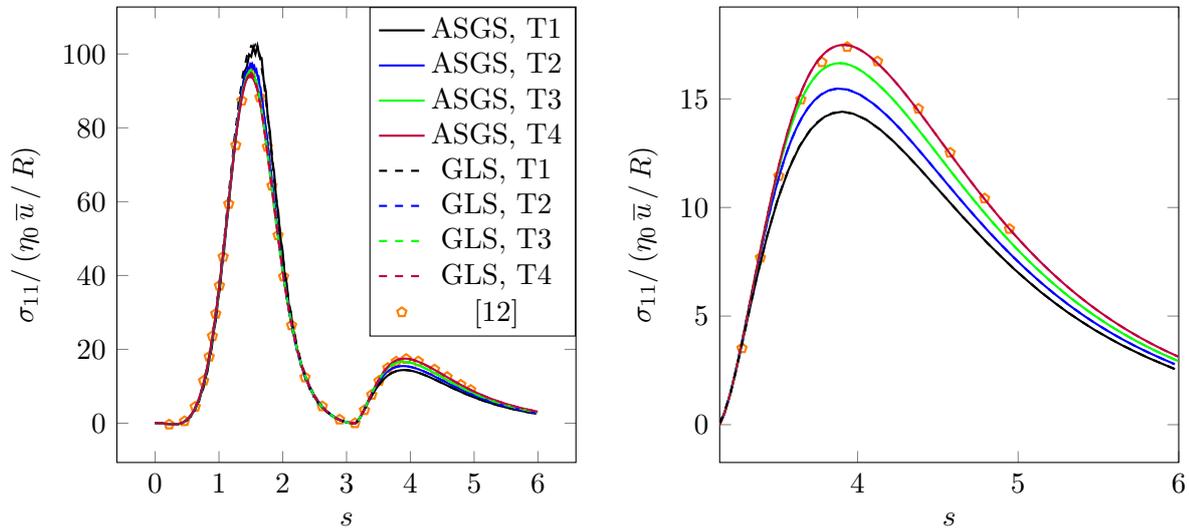

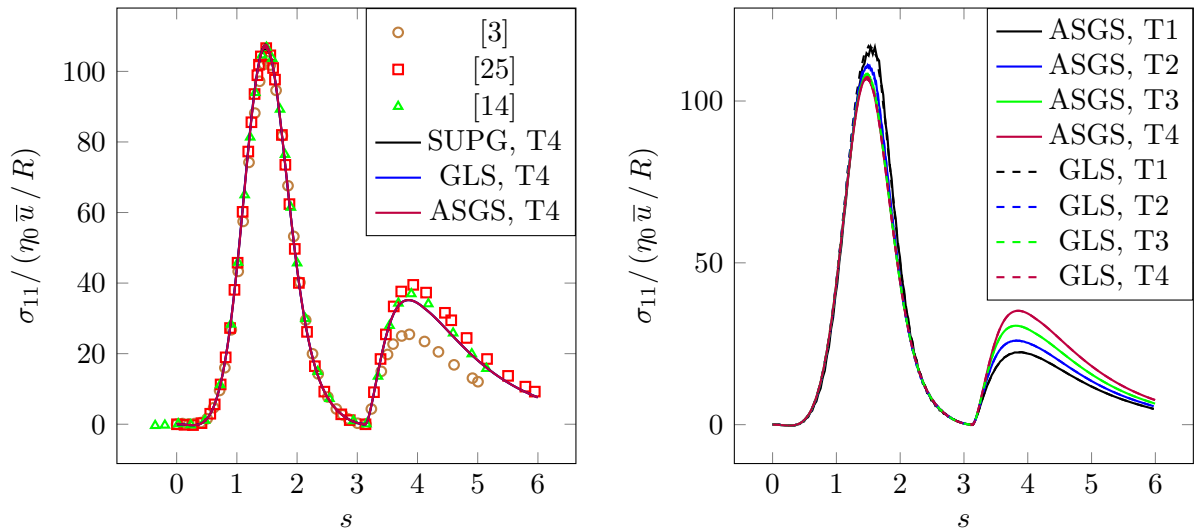
\begin{figure}[H]		
\subfigure{							
					\begin{tikzpicture}[scale=1.0]
					\begin{axis}
					[
					set layers,					
					mark layer=axis background,
					width=0.45\textwidth,
				    height=0.45\textwidth,
					xlabel=\small{$s$},
					ylabel=\small{$\sigma_{11}/ \br{\etaT \, \overline{u} \, / \, R}$},
					grid style=off,
					ticklabel style = {font=\small},
					legend style={at={(1.0,1.0)},anchor=north east,font=\small},
					xtick={0,1,2,3,4,5,6},
					]
					\addplot[color = brown, mark=o, mark options={scale=0.9}, only marks, thick] table [x=s, y=txx, col sep=comma] {fig/Cylinder2to1_Lineplots/References/Cylinder2to1_Lineplot_Alves2001_Wi0.7000.csv};\addlegendentry{\cite{Alves2001}};	
					
					\addplot[color = red, mark=square, mark options={scale=0.9}, only marks, thick] table [x=s, y=txx, col sep=comma] {fig/Cylinder2to1_Lineplots/References/Cylinder2to1_Lineplot_Knechtges2014_Wi0.7000.csv};\addlegendentry{\cite{Knechtges2014a}};	
					
					\addplot[color = green, mark=triangle, mark options={scale=0.9}, only marks, thick] table [x=s, y=txx, col sep=comma] {fig/Cylinder2to1_Lineplots/References/Cylinder2to1_Lineplot_Fan1999_Wi0.7000.csv};\addlegendentry{\cite{Fan99}};				
					
					\addplot[color = black, mark=none*, style=thick, solid] table [x=s, y=txx, col sep=comma] {fig/Cylinder2to1_Lineplots/Cylinder2to1_Lineplot_GLS2_T5_Wi0.7000.csv};\addlegendentry{SUPG, T4};	
					
					\addplot[color = blue, mark=none*, style=thick, solid] table [x=s, y=txx, col sep=comma] {fig/Cylinder2to1_Lineplots/Cylinder2to1_Lineplot_GLS3_T5_Wi0.7000.csv};\addlegendentry{GLS, T4};	
										
					\addplot[color = purple, mark=none*, style=thick, solid] table [x=s, y=txx, col sep=comma] {fig/Cylinder2to1_Lineplots/Cylinder2to1_Lineplot_ASS_T5_Wi0.7000.csv};\addlegendentry{ASGS, T4};		
					
					\end{axis}
					\end{tikzpicture}
}~
\subfigure{		
					\begin{tikzpicture}[scale=1.0]
					\begin{axis}
					[
					set layers,					
					mark layer=axis background,
					width=0.45\textwidth,
				    height=0.45\textwidth,
					xlabel=\small{$s$},
					ylabel=\small{$\sigma_{11}/ \br{\etaT \, \overline{u} \, / \, R}$},
					grid style=off,
					ticklabel style = {font=\small},
					legend style={at={(1.0,1.0)},anchor=north east,font=\small},
					xtick={0,1,2,3,4,5,6},
					]		
					
					\addplot[color = black, mark=none*, style=thick, solid] table [x=s, y=txx, col sep=comma] {fig/Cylinder2to1_Lineplots/Cylinder2to1_Lineplot_ASS_T2_Wi0.7000.csv};\addlegendentry{ASGS, T1};		
					
					\addplot[color = blue, mark=none*, style=thick, solid] table [x=s, y=txx, col sep=comma] {fig/Cylinder2to1_Lineplots/Cylinder2to1_Lineplot_ASS_T3_Wi0.7000.csv};\addlegendentry{ASGS, T2};		
					
					\addplot[color = green, mark=none*, style=thick, solid] table [x=s, y=txx, col sep=comma] {fig/Cylinder2to1_Lineplots/Cylinder2to1_Lineplot_ASS_T4_Wi0.7000.csv};\addlegendentry{ASGS, T3};		
					
					\addplot[color = purple, mark=none*, style=thick, solid] table [x=s, y=txx, col sep=comma] {fig/Cylinder2to1_Lineplots/Cylinder2to1_Lineplot_ASS_T5_Wi0.7000.csv};\addlegendentry{ASGS, T4};		
					
					\addplot[color = black, mark=none*, style=thick, dashed] table [x=s, y=txx, col sep=comma] {fig/Cylinder2to1_Lineplots/Cylinder2to1_Lineplot_GLS3_T2_Wi0.7000.csv};\addlegendentry{GLS, T1};		
					
					\addplot[color = blue, mark=none*, style=thick, dashed] table [x=s, y=txx, col sep=comma] {fig/Cylinder2to1_Lineplots/Cylinder2to1_Lineplot_GLS3_T3_Wi0.7000.csv};\addlegendentry{GLS, T2};		
					
					\addplot[color = green, mark=none*, style=thick, dashed] table [x=s, y=txx, col sep=comma] {fig/Cylinder2to1_Lineplots/Cylinder2to1_Lineplot_GLS3_T4_Wi0.7000.csv};\addlegendentry{GLS, T3};		
					
					\addplot[color = purple, mark=none*, style=thick, dashed] table [x=s, y=txx, col sep=comma] {fig/Cylinder2to1_Lineplots/Cylinder2to1_Lineplot_GLS3_T5_Wi0.7000.csv};\addlegendentry{GLS, T4};					
					
					\end{axis}
					\end{tikzpicture}
}
\caption{Stress profiles on the cylinder wall and wake at $\te{Wi}=0.7$}
\label{fig:cyllineplot07}
\end{figure}

We display the first normal stress component along a parametrized curve $s$ following the cylinder boundary and the wake for further validation purposes.
A comparison to values by Damanik \cite{Damanik2010} for Weissenberg number $\Wi=0.6$ in Fig.~\ref{fig:cyllineplotmeshes06} suggests being close to a converged profile.
Fig.~\ref{fig:cyllineplot07} compares our computations at $\Wi=0.7$ with results of Alves \cite{Alves2001}, Knechtges \cite{Knechtges2014a}, and Fan \cite{Fan99}. There is excellent agreement on the cylinder ($0\leq s \leq \pi$) for all methods, yet little to no agreement in the wake ($s\geq\pi$). Discrepancies in the wake starting at $\Wi=0.7$ are known, and the accurate prediction of the stresses in the wake is still a challenge for this benchmark \cite{Alves2021}. Fan \cite{Fan2005} questioned the existence of physical solutions beyond $\Wi\geq 0.8$ due to the stress boundary layers in the rear wake downstream the stagnation point. It was further conjectured that even if numerical solutions can be obtained for higher Weissenberg numbers, approximation errors might change the constitutive law locally. Renardy \cite{Renardy2006} has demonstrated for a square geometry that the extra stresses lose their smoothness with increasing Weissenberg number. The considered flow field shows a stagnation point from which downstream singularities in the extra stresses can be expected. A similar situation occurs in the rear wake of the cylinder.
					
\subsection{Flow through a 4:1 planar contraction}
\label{sec:4to1}

\begin{figure}[H]
\centering
\def\svgwidth{0.7\textwidth}
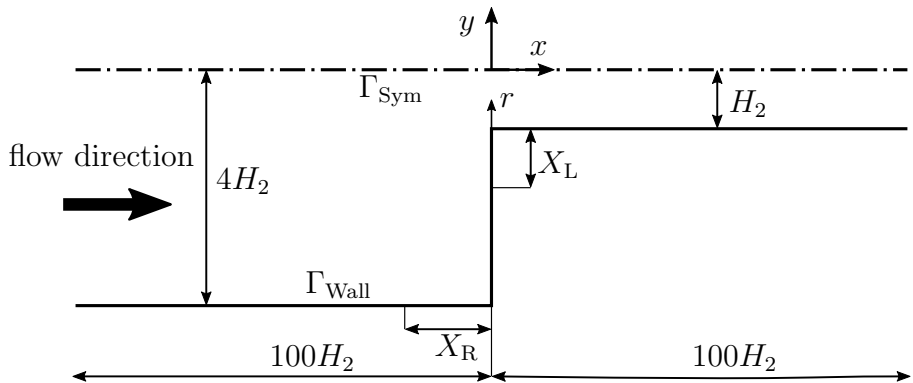
\caption{Schematic drawing of the flow through a 4:1 planar contraction.}
\label{fig:schematicContr4to1}
\end{figure}

The last benchmark presented is the flow through an abrupt planar contraction with a 4:1 ratio of the channel heights. Fig.~\ref{fig:schematicContr4to1} displays the geometry. Half the domain is computed; thus, symmetry conditions, $u_2=0$ and $\chi_{12}=0$, are imposed on the top boundary. The no-slip condition, $\vect{u}=\vect{0}$, is imposed on the bottom boundary. At the outlet, in addition to set $u_2=0$, we further prescribed a parabolic velocity profile $u_1 = \frac{3}{2} \overline{u}_2 \br{1 - \br{\frac{y}{H_2}}^2}$, and the pressure to zero. We have found these additional conditions useful to stabilize the flow field at the outlet at high Weissenberg numbers. The inlet conditions are
\begin{align}
u_1 &= \frac{3}{8} \overline{u}_2 \br{1 - \br{\frac{y}{4H_2}}^2} \, ,\, u_2 = 0\,,\\
\sigma_{11} &= \frac{9}{8} \lambda (1-\beta) \etaT \overline{u}_2^2 \frac{y^2}{\br{4H_2}^4} \, ,\, \sigma_{12} = -\frac{3}{4} (1-\beta) \etaT \overline{u}_2 \frac{y}{\br{4H_2}^2} \, ,\, \te{and} \,\sigma_{22} = 0 \,,
\end{align}
which are transformed to $\ten{\chi}$ as described in Sec.~\ref{sec:cylinder}. The parameters for this benchmark are $\beta=1/9$, $\etaT=1$, $H_2=1$, and the outlet mean velocity is set to $\overline{u}_2=1$. The Weissenberg number is defined as $\Wi = \lambda \overline{u}_2/H_2$.

This flow configuration is considered a very stringent benchmark due to the unbounded behavior of the pressure and extra stresses near the re-entrant corner situated at $(x,y)=(0,-H_2)$ for increasingly elastic flows. Characteristic for the flow field in the 4:1 contraction is the corner vortex in the wider channel and the lip vortex near the re-entrant corner. The sizes of the corner and lip vortex are $X_{\te{R}}$ and $X_{\te{L}}$, respectively. The lip vortex emerges due to normal stresses near the re-entrant corner and can not be observed in a Newtonian flow. The size of the corner vortex is defined as the upstream distance of the reattachment point to the contraction plane. The reattachment point is located, where the z-component of the vorticity is zero, i.e., $\omega_{\te{wall}} =\br{\partial_{x_1} u_2 - \partial_{x_2} u_1} \bigg\rvert_{\te{wall}} = 0$. The size of the lip vortex is defined by the distance between the reattachment point on the vertical wall to the re-entrant corner.

\begin{figure}[H]
\includegraphics[trim=0 0 0 0, clip, width=1.0\textwidth]{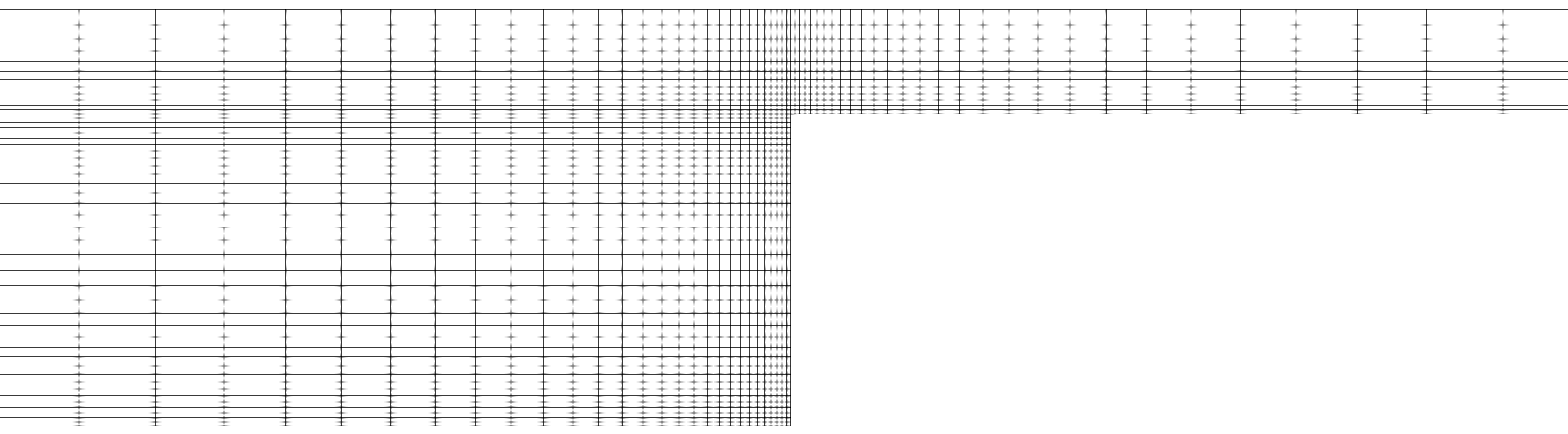}
\caption{Quadrilateral mesh Q1 for the flow through a 4:1 planar contraction}
\label{fig:4to1mesh}
\end{figure}

Tab.~\ref{tab:4to1mesh} collects relevant numbers of the employed computational meshes. 
We use quadrilateral meshes with successive element-stretching to refine near the re-entrant corner and near the lower wall (Fig.~\ref{fig:4to1mesh}). In our initial mesh designs, the upstream channel length revealed slight effects on the vortical structures near the contraction plane. Taking the up-and down-stream channel lengths a hundred times larger than the downstream channel height was sufficient to ensure fully developed profiles in both channels.
A pure element stretching strategy in the whole domain leads to extreme aspect ratios of elements near the in- and outlet. 
Numerical experiments suggested that meshes containing elements with aspect ratios larger than one hundred began to increase the required iterative steps in the linear solver. 
The meshes are uniform in $x$-direction for $x<-40$ and $x>40$ to limit the aspect ratios to about a hundred.
The tolerance of the non-linear iteration is set to $\epsilon=10^{-9}$.

\begin{table}[H]
\centering
\caption{Mesh characteristics for the flow through a 4:1 planar contraction}
\label{tab:4to1mesh}
\begin{tabular}{c c c c}
\hline
Mesh & Number of elements & Number of nodes & Smallest element near re-entrant corner \\
\hline
Q1 & 3480 & 3645 & 0.04 \\
Q2 & 14160 & 14493 & 0.02 \\
Q3 & 56640 & 57305 & 0.01 \\
\hline
\end{tabular}
\end{table}

\subsubsection{On limiting Weissenberg numbers}

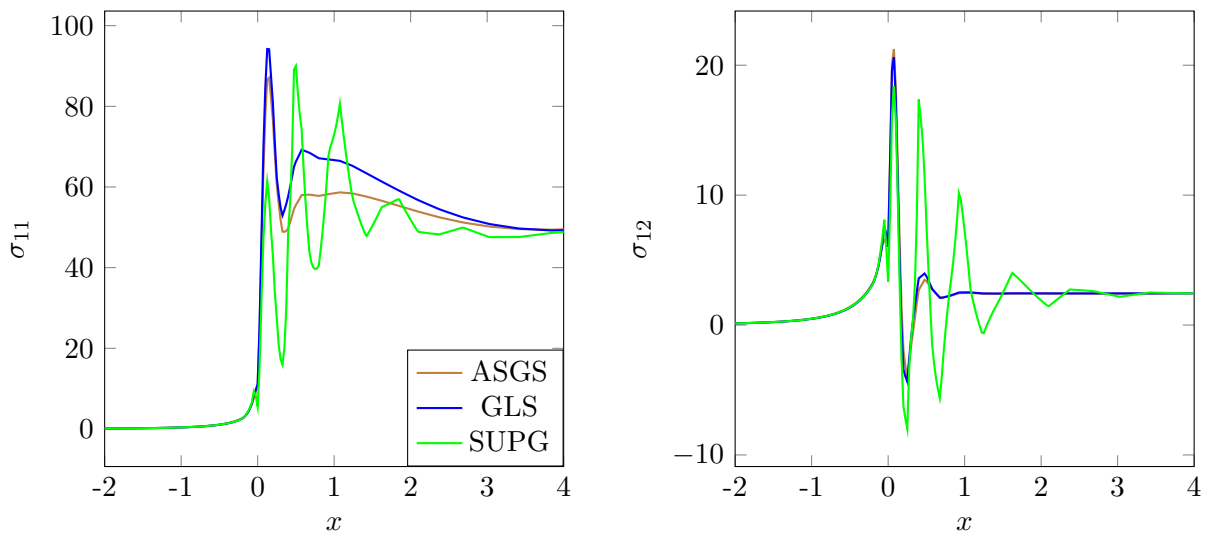
\begin{figure}[H]
	\subfigure{
	\begin{tikzpicture}
					\begin{axis}
					[
					set layers,					
					mark layer=axis background,
 					width=0.45\textwidth,
				    height=0.45\textwidth,
					xlabel=\small{$x$},
					ylabel=\small{$\sigma_{11}$},
					grid style=off,
					ticklabel style = {font=\small},
					legend style={at={(1.0,0.0)},anchor=south east,font=\small},
					xmin=98,
					xmax=104,
					xtick={98, 99, 100, 101, 102, 103, 104},
					xticklabels={-2, -1, 0, 1, 2, 3, 4}
					]
					\addplot[color = brown, mark=none*, thick, solid] table [x=x, y=t11, col sep=comma] {fig/FourToOneContraction/CornerLineData/CornerLineData.formASS.lambda4.0000.Ref1.csv};\addlegendentry{ASGS};
					
					\addplot[color = blue, mark=none*, thick, solid] table [x=x, y=t11, col sep=comma] {fig/FourToOneContraction/CornerLineData/CornerLineData.formGLS3.lambda4.0000.Ref1.csv};\addlegendentry{GLS};

					\addplot[color = green, mark=none*, thick, solid] table [x=x, y=t11, col sep=comma] {fig/FourToOneContraction/CornerLineData/CornerLineData.formGLS2.lambda4.0000.Ref1.csv};\addlegendentry{SUPG};
					\end{axis}
	\end{tikzpicture}
	}~
	\subfigure{
	\begin{tikzpicture}
					\begin{axis}
					[
					set layers,					
					mark layer=axis background,
					width=0.45\textwidth,
				    height=0.45\textwidth,
					xlabel=\small{$x$},
					ylabel=\small{$\sigma_{12}$},
					grid style=off,
					ticklabel style = {font=\small},
					legend style={at={(0.9,0.2)},anchor=north west,font=\small},
					xmin=98,
					xmax=104,
					xtick={98, 99, 100, 101, 102, 103, 104},
					xticklabels={-2, -1, 0, 1, 2, 3, 4}
					]
					\addplot[color = brown, mark=none*, thick, solid] table [x=x, y=t12, col sep=comma] {fig/FourToOneContraction/CornerLineData/CornerLineData.formASS.lambda4.0000.Ref1.csv};
					
					\addplot[color = blue, mark=none*, thick, solid] table [x=x, y=t12, col sep=comma] {fig/FourToOneContraction/CornerLineData/CornerLineData.formGLS3.lambda4.0000.Ref1.csv};

					\addplot[color = green, mark=none*, thick, solid] table [x=x, y=t12, col sep=comma] {fig/FourToOneContraction/CornerLineData/CornerLineData.formGLS2.lambda4.0000.Ref1.csv};
					\end{axis}
	\end{tikzpicture}
	}
\caption{Stress profiles along $y=-H_2$ for mesh Q1 at $\Wi=4.0$.}
\label{fig:oscillations}
\end{figure}

Tab.~\ref {tab:4to1limwi} summarizes our limiting Weissenberg numbers at which the numerical computation breaks down. Significantly lower limiting numbers for the SUPG method are observed. Fig.~\ref{fig:oscillations} displays the first normal stress component and off-diagonal stress component along the line $y=-H_2$ for our coarsest mesh. Whereas the GLS and ASGS method reveal a similar graph, the SUPG is polluted with oscillations downstream of the re-entrant corner. The SUPG method does not introduce an elliptic contribution, which would allow the bypass of the stress-velocity compatibility condition. The higher solvent viscosity ratio in the cylinder benchmark possibly explains why we could not detect significant differences between the SUPG and GLS approach. In the present case, the geometric singularity generates high-velocity gradients that drive viscoelastic stresses. The need for a method that does provide control on the velocity gradients is substantiated by the observed poor performance of the SUPG approach. In the following, the SUPG method will not be further considered in our comparisons.

The computations on our most refined mesh with the GLS method were computed on a single core, as the iterative FGMRES method was no longer able to converge for parallel computations. In the case of the ASGS approach, 12 cores could be used on every presented mesh without any difficulty. However, the element size near the re-entrant corner posed clear limitation to our computations. Finer resolutions led to an early breakdown at $\Wi= 0.8$ at most. Lower limiting Weissenberg numbers on finer mesh resolutions is a primary characteristic of the HWNP \cite{Keunings1986}. This benchmark seems to illustrate clearly that the HWNP is alleviated with a logarithmic reformulation but not eliminated.

\begin{table}[H]
\centering
\caption{Limiting Weissenberg numbers for the flow through a 4:1 planar contraction of presented methods and meshes.}
\label{tab:4to1limwi}
\begin{tabular}{l|llll}
                     & Q1   & Q2   & Q3  \\
\hline
SUPG                 & 8.27 & 2.365 & 1.635\\
GLS                  & 15.18 & 13.53 & 3.4\\
ASGS                 & 15.77 & 15.47 & 14.45
\end{tabular}
\end{table}

\subsubsection{Asymptotic slopes of stress and velocity near re-entrant corner}

\begin{figure}[H]
	\subfigure[]{
	\begin{tikzpicture}
					\begin{axis}
					[
					set layers,					
					mark layer=axis background,
					width=0.45\textwidth,
				    height=0.55\textwidth,
					xlabel=\small{$\log{\br{r/H_2}}$},
					ylabel=\small{$\log{\br{\abs{\sigma_{ij}}/\br{\etaT \overline{u} / H_2}}}$},
					grid style=off,
					ticklabel style = {font=\small},
					legend style={at={(0.0,0.0)},anchor=south west,font=\small},
					xmin=-5,
					xmax=0,
					]
					\addplot[color= orange, mark=square, mark options={scale=1.0}, only marks, thick] table [x=r, y=t11, col sep=comma] {fig/FourToOneContraction/AsymptoticSlopes/AsymptoticData.formASS.Wi1.0000.Ref1.csv};\addlegendentry{Q1, $\sigma_{11}$};		
					\addplot[color= blue, mark=square, mark options={scale=1.0}, only marks, thick] table [x=r, y=t11, col sep=comma] {fig/FourToOneContraction/AsymptoticSlopes/AsymptoticData.formASS.Wi1.0000.Ref2.csv};\addlegendentry{Q2, $\sigma_{11}$};										
					\addplot[color= red, mark=square, mark options={scale=1.0}, only marks, thick] table [x=r, y=t11, col sep=comma] {fig/FourToOneContraction/AsymptoticSlopes/AsymptoticData.formASS.Wi1.0000.Ref3.csv};\addlegendentry{Q3, $\sigma_{11}$};					
					
					\addplot[color= orange, mark=o, mark options={scale=1.0}, only marks, thick] table [x=r, y=t12, col sep=comma] {fig/FourToOneContraction/AsymptoticSlopes/AsymptoticData.formASS.Wi1.0000.Ref1.csv};\addlegendentry{Q1, $\sigma_{12}$};		
					\addplot[color= blue, mark=o, mark options={scale=1.0}, only marks, thick] table [x=r, y=t12, col sep=comma] {fig/FourToOneContraction/AsymptoticSlopes/AsymptoticData.formASS.Wi1.0000.Ref2.csv};\addlegendentry{Q2, $\sigma_{12}$};					
					\addplot[color= red, mark=o, mark options={scale=1.0}, only marks, thick] table [x=r, y=t12, col sep=comma] {fig/FourToOneContraction/AsymptoticSlopes/AsymptoticData.formASS.Wi1.0000.Ref3.csv};\addlegendentry{Q3, $\sigma_{12}$};

					\addplot[color= orange, mark=pentagon, mark options={scale=1.0}, only marks, thick] table [x=r, y=t22, col sep=comma] {fig/FourToOneContraction/AsymptoticSlopes/AsymptoticData.formASS.Wi1.0000.Ref1.csv};\addlegendentry{Q1, $\sigma_{22}$};					
					\addplot[color= blue, mark=pentagon, mark options={scale=1.0}, only marks, thick] table [x=r, y=t22, col sep=comma] {fig/FourToOneContraction/AsymptoticSlopes/AsymptoticData.formASS.Wi1.0000.Ref2.csv};\addlegendentry{Q2, $\sigma_{22}$};
					\addplot[color= red, mark=pentagon, mark options={scale=1.0}, only marks, thick] table [x=r, y=t22, col sep=comma] {fig/FourToOneContraction/AsymptoticSlopes/AsymptoticData.formASS.Wi1.0000.Ref3.csv};\addlegendentry{Q3, $\sigma_{22}$};
					
					\addplot[color= black, mark=none*, solid] table [x=r, y=t11ref, col sep=comma] {fig/FourToOneContraction/AsymptoticSlopes/AsymptoticData.formASS.Wi1.0000.Ref3.csv};
					\addplot[color= black, mark=none*, solid] table [x=r, y=t12ref, col sep=comma] {fig/FourToOneContraction/AsymptoticSlopes/AsymptoticData.formASS.Wi1.0000.Ref3.csv};
					\addplot[color= black, mark=none*, solid] table [x=r, y=t22ref, col sep=comma] {fig/FourToOneContraction/AsymptoticSlopes/AsymptoticData.formASS.Wi1.0000.Ref3.csv};

					\end{axis}
	\end{tikzpicture}
	}~
	\subfigure[]{
	\begin{tikzpicture}
					\begin{axis}
					[
					set layers,					
					mark layer=axis background,
					width=0.45\textwidth,
				    height=0.55\textwidth,
					xlabel=\small{$\log{\br{r/H_2}}$},
					ylabel=\small{$\log{\br{\abs{u_i}/ \overline{u}}}$},
					grid style=off,
					ticklabel style = {font=\small},
					legend style={at={(0.3,0.0)},anchor=south west,font=\small},
					xmin=-5,
					xmax=0,
					]
					\addplot[color= orange, mark=square, mark options={scale=1.0}, only marks, thick] table [x=r, y=u1, col sep=comma] {fig/FourToOneContraction/AsymptoticSlopes/AsymptoticData.formASS.Wi1.0000.Ref1.csv};\addlegendentry{Q1, $u_1$};	
					\addplot[color= blue, mark=square, mark options={scale=1.0}, only marks, thick] table [x=r, y=u1, col sep=comma] {fig/FourToOneContraction/AsymptoticSlopes/AsymptoticData.formASS.Wi1.0000.Ref2.csv};\addlegendentry{Q2, $u_1$};	
					\addplot[color= red, mark=square, mark options={scale=1.0}, only marks, thick] table [x=r, y=u1, col sep=comma] {fig/FourToOneContraction/AsymptoticSlopes/AsymptoticData.formASS.Wi1.0000.Ref3.csv};\addlegendentry{Q3, $u_1$};				

					\addplot[color= orange, mark=o, mark options={scale=1.0}, only marks, thick] table [x=r, y=u2, col sep=comma] {fig/FourToOneContraction/AsymptoticSlopes/AsymptoticData.formASS.Wi1.0000.Ref1.csv};\addlegendentry{Q1, $u_2$};					
					\addplot[color= blue, mark=o, mark options={scale=1.0}, only marks, thick] table [x=r, y=u2, col sep=comma] {fig/FourToOneContraction/AsymptoticSlopes/AsymptoticData.formASS.Wi1.0000.Ref2.csv};\addlegendentry{Q2, $u_2$};
					\addplot[color= red, mark=o, mark options={scale=1.0}, only marks, thick] table [x=r, y=u2, col sep=comma] {fig/FourToOneContraction/AsymptoticSlopes/AsymptoticData.formASS.Wi1.0000.Ref3.csv};\addlegendentry{Q3, $u_2$};

					\addplot[color= black, mark=none*, solid] table [x=r, y=u1ref, col sep=comma] {fig/FourToOneContraction/AsymptoticSlopes/AsymptoticData.formASS.Wi1.0000.Ref3.csv};
					\addplot[color= black, mark=none*, solid] table [x=r, y=u2ref, col sep=comma] {fig/FourToOneContraction/AsymptoticSlopes/AsymptoticData.formASS.Wi1.0000.Ref3.csv};

					\end{axis}
	\end{tikzpicture}
	}
					\caption{Comparison of stress profiles with theoretical slope of $-2/3$ (a), and velocity profiles with theoretical slope of $5/9$ (b) along $x=0$ for $\te{Wi}=1.0$ computed with the ASGS method.}
					\label{fig:4to1asym}
\end{figure}
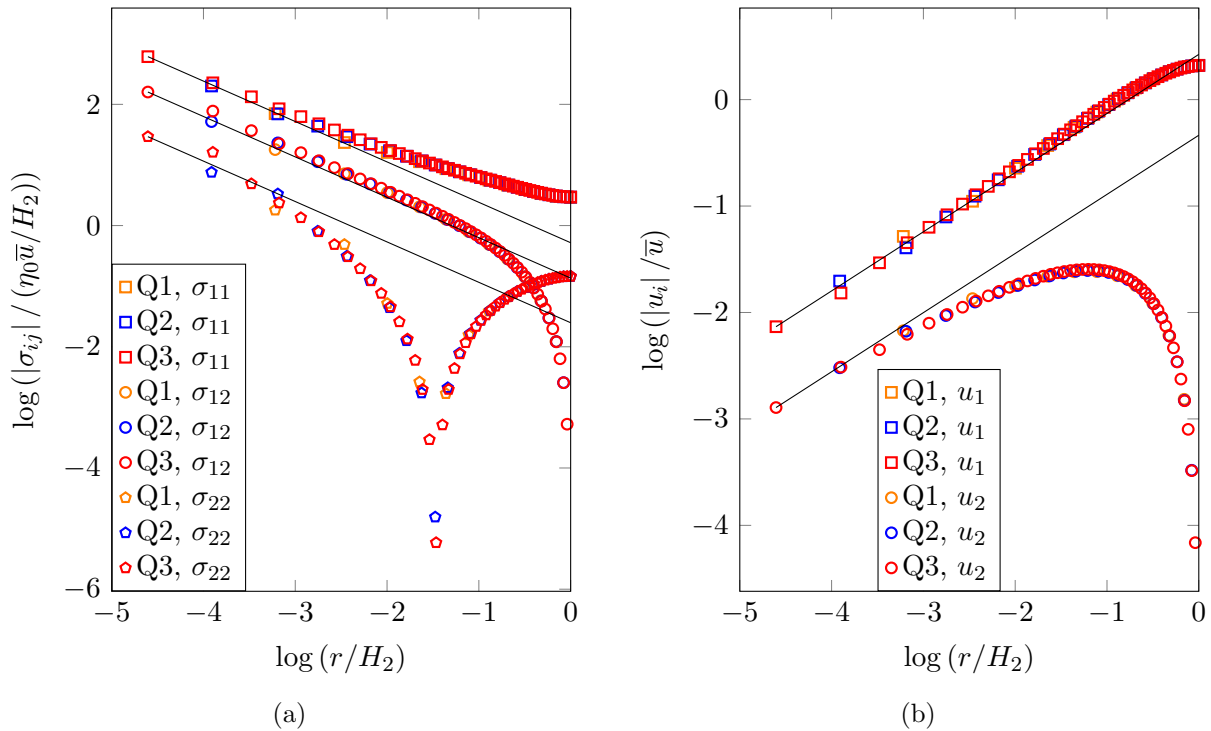

We proceed with a validation study for a low Weissenberg number $\Wi=1$. Before the onset of the lip vortex, the stress and velocity can be described asymptotically in the vicinity of the re-entrant corner. Evans \cite{Evans2005} postulated the stresses behave as $\sigma_{ij}\propto r^{-2/3}$ and the velocity as $u_i \propto r^{5/9}$ near the singular point, where $r$ is the distance to the re-entrant corner across the contraction plane (see Fig.~\ref{fig:schematicContr4to1}). Fig.~\ref{fig:4to1asym} compares the profiles for stresses and velocity of the ASGS method for different meshes to the theoretically predicted slopes. Qualitatively, we observe similar behavior of stress and velocity components as shown in \cite{Pimenta2017}. A surprisingly little mesh-dependence is observed, indicating converged profiles. Close to the singular point, a good agreement to the theoretical predictions can be observed. The deviations become larger with increasing distance to the re-entrant corner. The components $\sigma_{12}$ and $u_1$ reveal the closest match to theoretical slopes.

\subsubsection{Prediction of corner vortex sizes}
\label{sec:vortexsizes}

The most widely used parameter to assess the accuracy of new methods is the size of the corner vortex as a function of the Weissenberg number. Even in the low Weissenberg number regime ($\Wi \leq 3$), discrepancies among the different works remained. Agreement between the data could be found in 2003 and following years by works of, e.g., \cite{Alves2003,Castillo2014,Comminal2016} up to $\Wi=3$, which used to be the critical Weissenberg number for this benchmark problem. Logarithmic reformulations allowed the exploration of flows above the critical number. As basis for our comparison, we refer to works with the highest reported Weissenberg numbers known by the authors \cite{Castillo2014,Moreno2019,Afonso2011,Comminal2016,Pimenta2017,Niethammer2018}.

\begin{figure}[H]
\subfigure[$\Wi=3$]{
\includegraphics[trim=8cm 0 30cm 0, clip, width=0.4\textwidth]{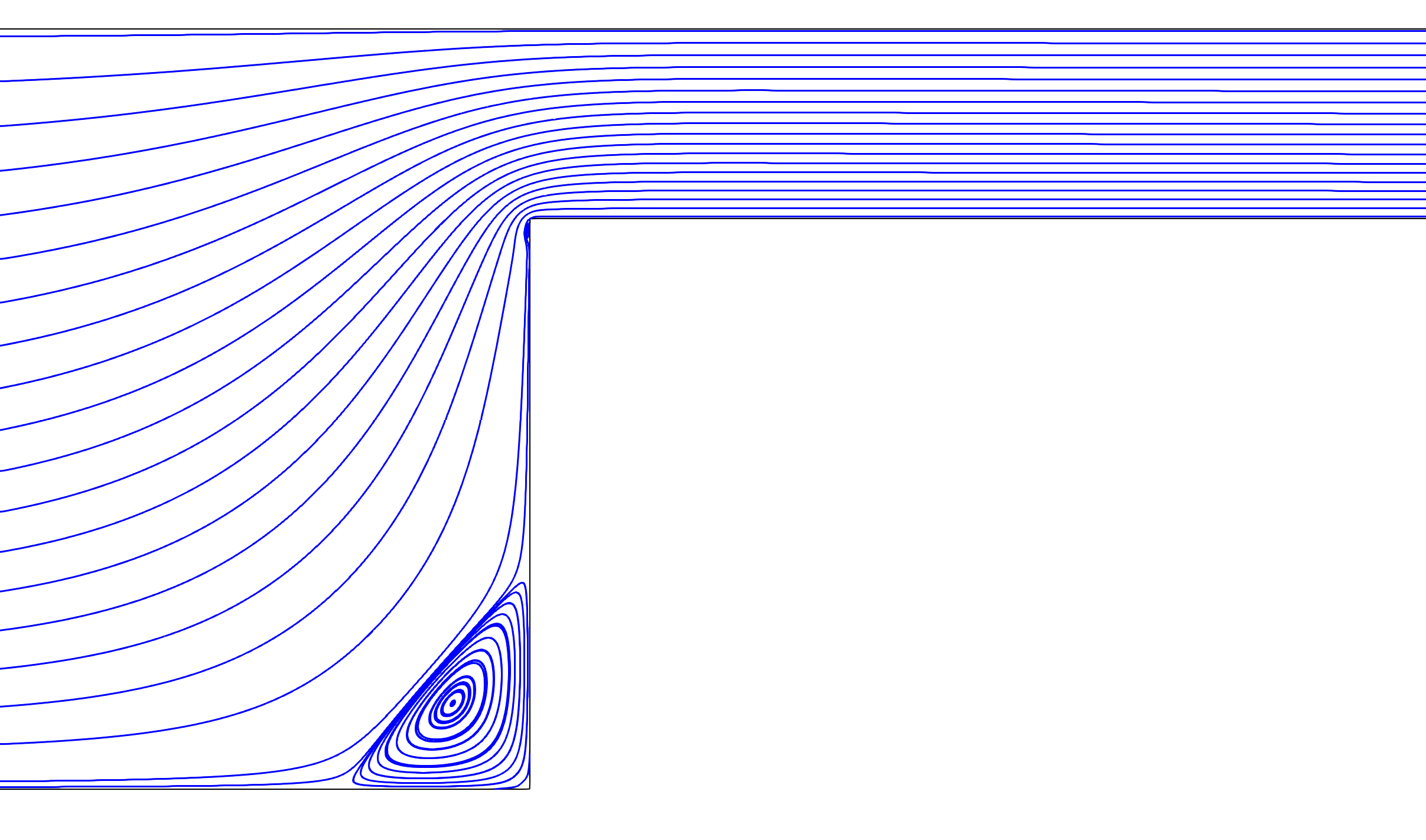}
}~
\subfigure[$\Wi=6$]{
\includegraphics[trim=8cm 0 30cm 0, clip, width=0.4\textwidth]{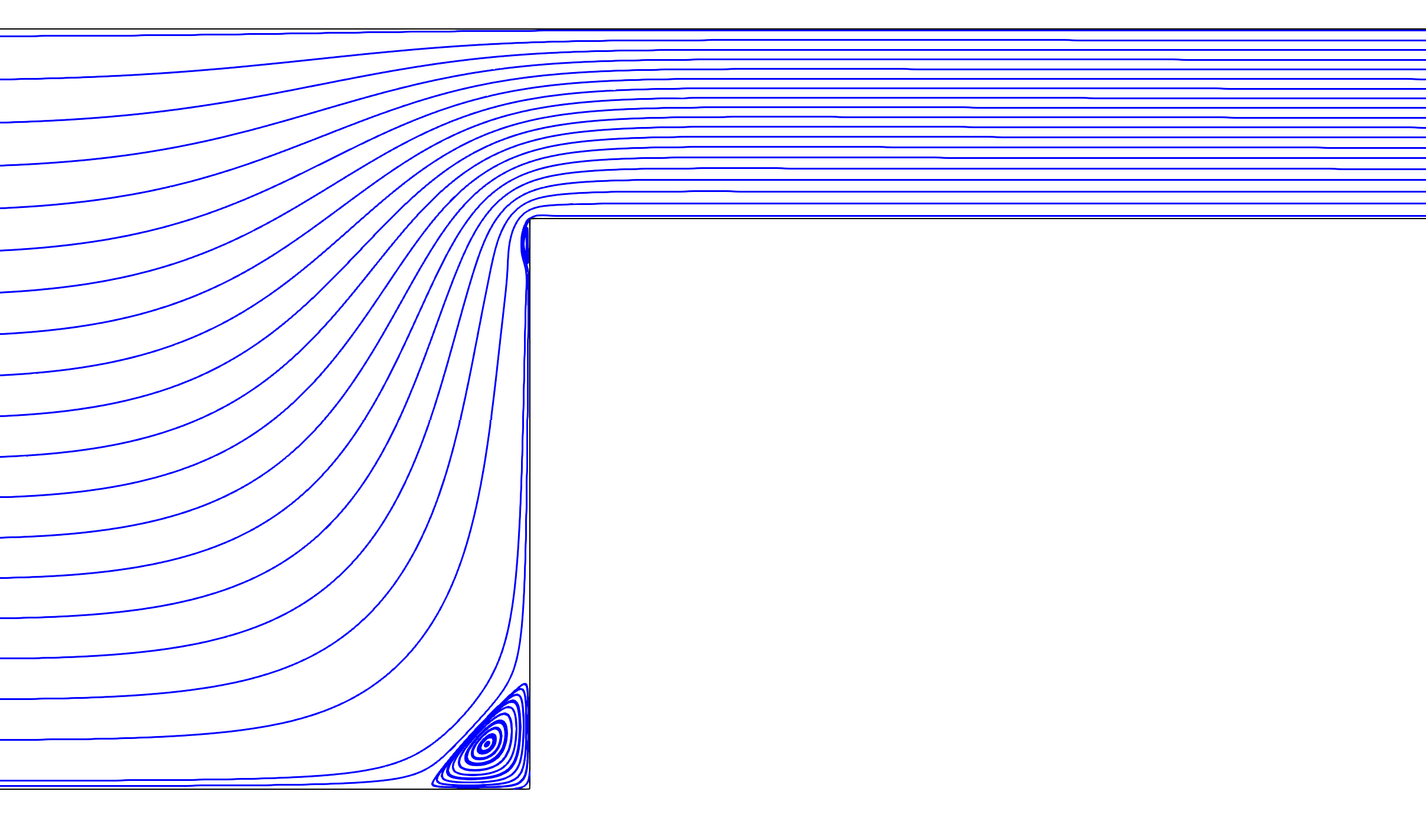}
}\\
\subfigure[$\Wi=9$]{
\includegraphics[trim=8cm 0 30cm 0, clip, width=0.4\textwidth]{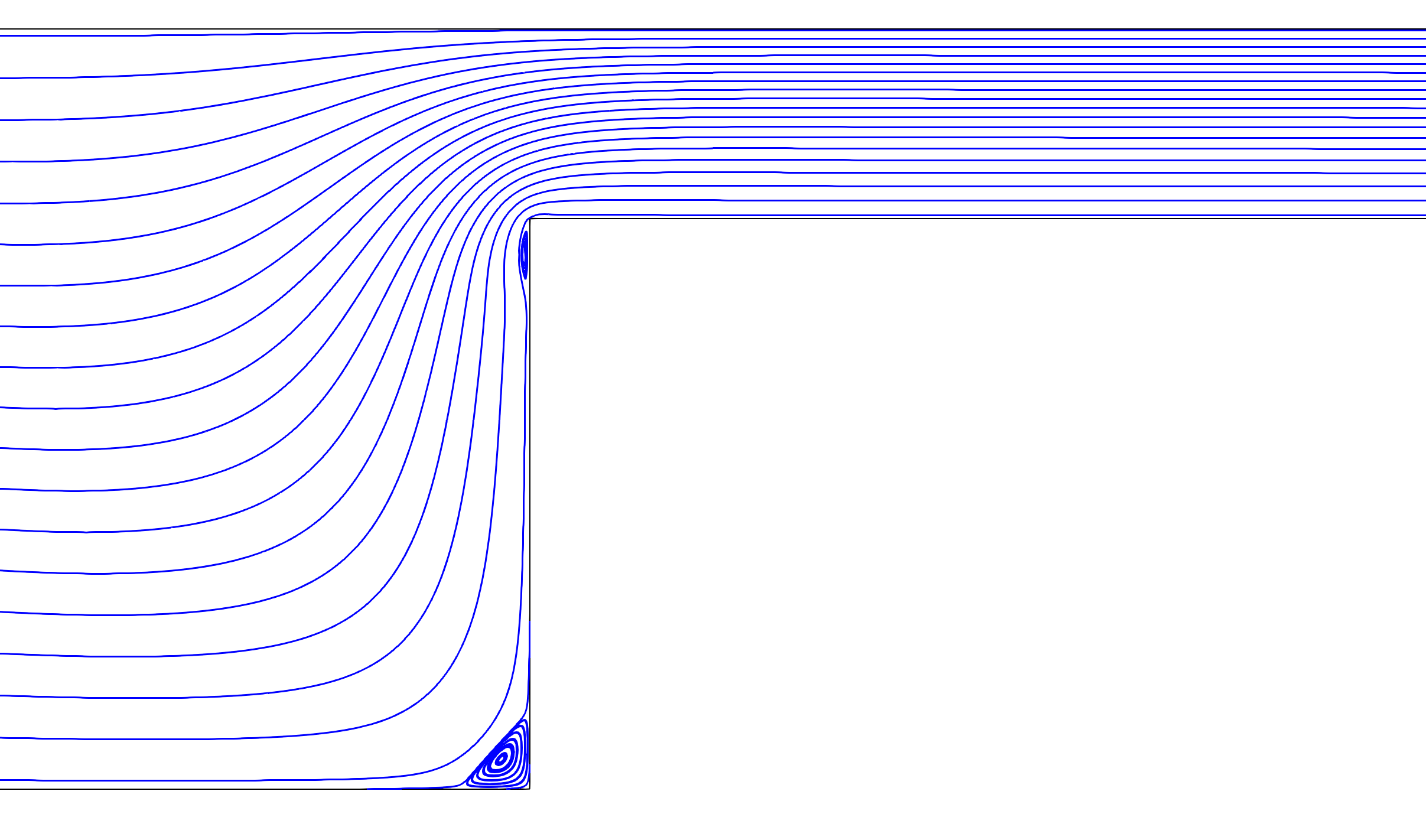}
}~
\subfigure[$\Wi=12$]{
\includegraphics[trim=8cm 0 30cm 0, clip, width=0.4\textwidth]{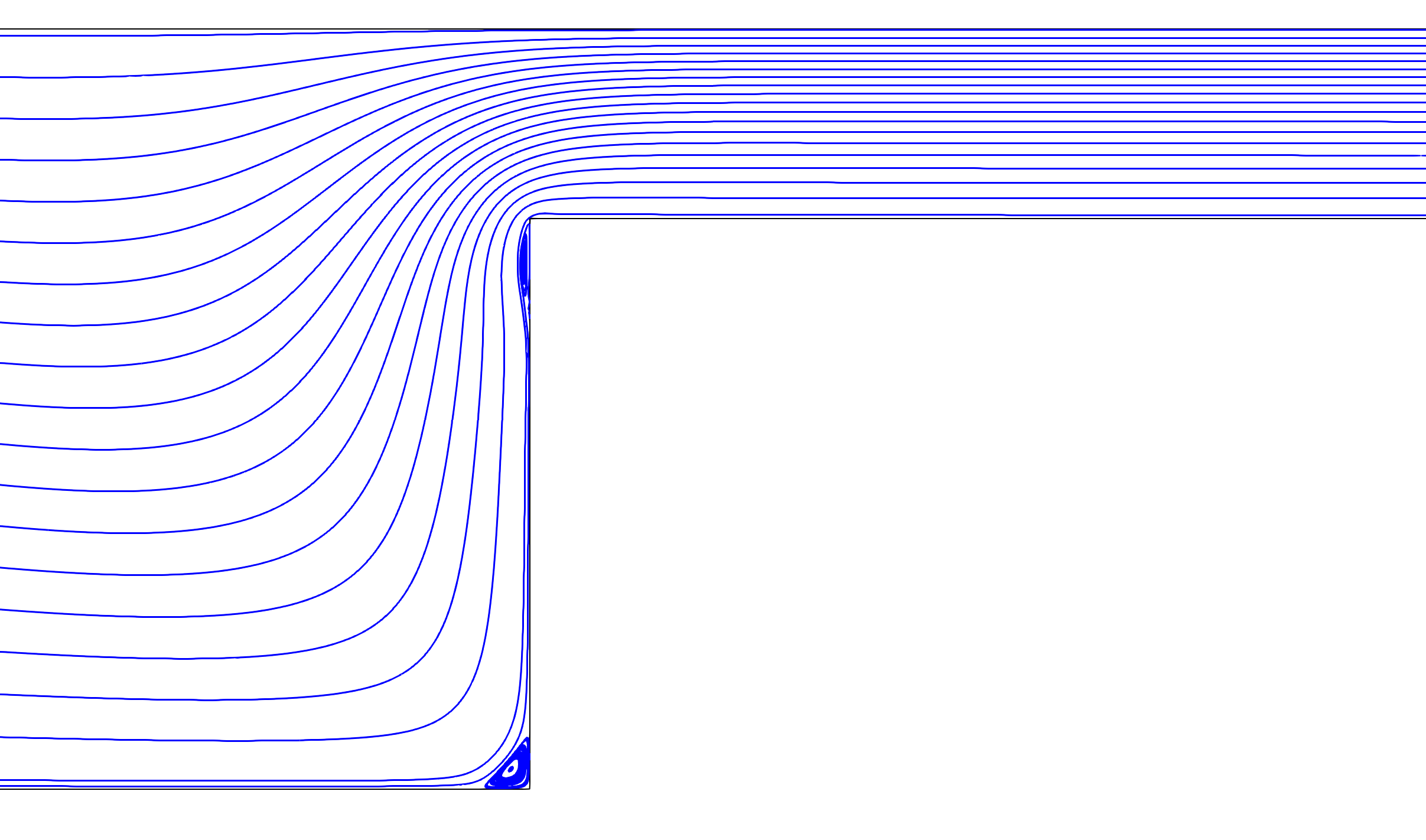}
}
\caption{Velocity streamlines of the flow through a 4:1 planar contraction with the ASGS method.}
\label{fig:streamlines4to1}
\end{figure}

Fig.~\ref{fig:4to1cornervortexlit} and Tab.~\ref{tab:4to1cornervortexlit} compares our findings with literature data. Our results are in good agreement with the data of \cite{Niethammer2018} up $\Wi=9$, and with the data of \cite{Pimenta2017} up to $\Wi=12$. These two works were the first to present independently matching benchmark data for Weissenberg numbers well beyond the critical value, as pointed out in \cite{Alves2021}. Our data seems to reconfirm a decreasing corner vortex up to at least $\Wi=9$. In the reference values of \cite{Moreno2019}, \cite{Afonso2011}, and \cite{Niethammer2018}, the corner vortex begins to grow at $\Wi\approx4$ or $\Wi=10$. The reason for the growth is a merging of the corner vortex and lip vortex to a single larger vortex. None of the presented computations exhibited a lip vortex sufficiently large to merge with the corner vortex. Fig.~\ref{fig:streamlines4to1} presents the streamlines for different Weissenberg numbers near the contraction area. The lip vortex grows with the Weissenberg number, whereas the corner vortex shrinks. However, the lack of accuracy of logarithmic reformulations at high Weissenberg numbers --- which is reported in most works --- does not allow a conclusion whether the vortices merge eventually or not. In agreement with the literature, Fig.~\ref{fig:4to1lipvortexlit} reveals that coarser mesh resolutions produce larger lip vortices, i.e., a strongly mesh-dependent prediction. In comparison to, e.g., \cite{Castillo2014} and \cite{Pimenta2017}, we observe overall smaller lip vortices in our results. Our findings seem to further support that corner and lip vortex merely connect erroneously for $\Wi<9$, due to overly large predictions of the lip vortex.

Regarding the quality of the numerical results, the ASGS method is superior to the GLS method. Node-to-node oscillations are clearly visible for Weissenberg numbers greater than $\Wi \approx 10$ with the GLS method on the coarsest mesh (not shown), where the ASGS method produced a smooth solution. This instability also affected the prediction of the cortex vortex sizes, as can be seen in Fig.~\ref{fig:4to1cornervortexlit}.

\begin{figure}[H]
	\begin{tikzpicture}
					\begin{axis}
					[
					set layers,					
					mark layer=axis background,
					width=0.7\textwidth,
				    height=0.45\textwidth,
					xlabel=\small{$\te{Wi}$},
					ylabel=\small{$X_{\te{R}}$},
					grid style=off,
					ticklabel style = {font=\small},
					legend style={at={(1.05,1.0)},anchor=north west,font=\small},
					xmin=0,
					xmax=16.0,
					xtick={0, 1.0, 2.0, 3.0, 4.0, 5.0, 6.0, 7.0, 8.0, 9.0, 10.0, 11.0, 12.0, 14.0, 16},
					]
					\addplot[color = blue, mark=none*, thick, dotted] table [x=Wi, y=VortexSize, col sep=comma] {fig/FourToOneContraction/VortexSize/CornerVortexSize_GLS3_Q1.csv};\addlegendentry{GLS, Q1};
					\addplot[color = blue, mark=none*, thick, dashed] table [x=Wi, y=VortexSize, col sep=comma] {fig/FourToOneContraction/VortexSize/CornerVortexSize_GLS3_Q2.csv};\addlegendentry{GLS, Q2};
					\addplot[color = blue, mark=none*, thick, solid] table [x=Wi, y=VortexSize, col sep=comma] {fig/FourToOneContraction/VortexSize/CornerVortexSize_GLS3_Q3.csv};\addlegendentry{GLS, Q3};

					\addplot[color = orange, mark=none*, thick, dotted] table [x=Wi, y=VortexSize, col sep=comma] {fig/FourToOneContraction/VortexSize/CornerVortexSize_ASS_Q1.csv};\addlegendentry{ASGS, Q1};
					\addplot[color = orange, mark=none*, thick, dashed] table [x=Wi, y=VortexSize, col sep=comma] {fig/FourToOneContraction/VortexSize/CornerVortexSize_ASS_Q2.csv};\addlegendentry{ASGS, Q2};
					\addplot[color = orange, mark=none*, thick, solid] table [x=Wi, y=VortexSize, col sep=comma] {fig/FourToOneContraction/VortexSize/CornerVortexSize_ASS_Q3.csv};\addlegendentry{ASGS, Q3};

					\addplot[color = green, mark=o, mark options={scale=1.0}, only marks, thick] table [x=Wi, y=VortexSize, col sep=comma] {fig/FourToOneContraction/VortexSizeReference/VortexSize_Alves.csv};\addlegendentry{Alves et al.\cite{Alves2003}};
					
					\addplot[color = red, mark=square, mark options={scale=1.0}, only marks, thick] table [x=Wi, y=VortexSize, col sep=comma] {fig/FourToOneContraction/VortexSizeReference/VortexSize_Pimenta.csv};\addlegendentry{Pimenta et al. \cite{Pimenta2017}};
					
					\addplot[color = purple, mark=pentagon, mark options={scale=1.0}, only marks, thick] table [x=Wi, y=VortexSize, col sep=comma] {fig/FourToOneContraction/VortexSizeReference/VortexSize_Niethammer.csv};\addlegendentry{Niethammer et al. \cite{Niethammer2018}};
					
					\addplot[color = orange, mark=star, mark options={scale=1.0}, only marks, thick] table [x=Wi, y=VortexSize, col sep=comma] {fig/FourToOneContraction/VortexSizeReference/VortexSize_Afonso.csv};\addlegendentry{Afonso et al. \cite{Afonso2011}};

					\addplot[color = pink, mark=diamond, mark options={scale=1.0}, only marks, thick] table [x=Wi, y=VortexSize, col sep=comma] {fig/FourToOneContraction/VortexSizeReference/VortexSize_Castillo.csv};\addlegendentry{Castillo et al. \cite{Castillo2014}};
					
					\addplot[color = brown, mark=triangle, mark options={scale=1.0}, only marks, thick] table [x=Wi, y=VortexSize, col sep=comma] {fig/FourToOneContraction/VortexSizeReference/VortexSize_Moreno.csv};\addlegendentry{Moreno et al. \cite{Moreno2019}};
					
					\end{axis}
					\end{tikzpicture}
\caption{Corner vortex sizes for different methods and mesh resolutions.}
\label{fig:4to1cornervortexlit}
\end{figure}
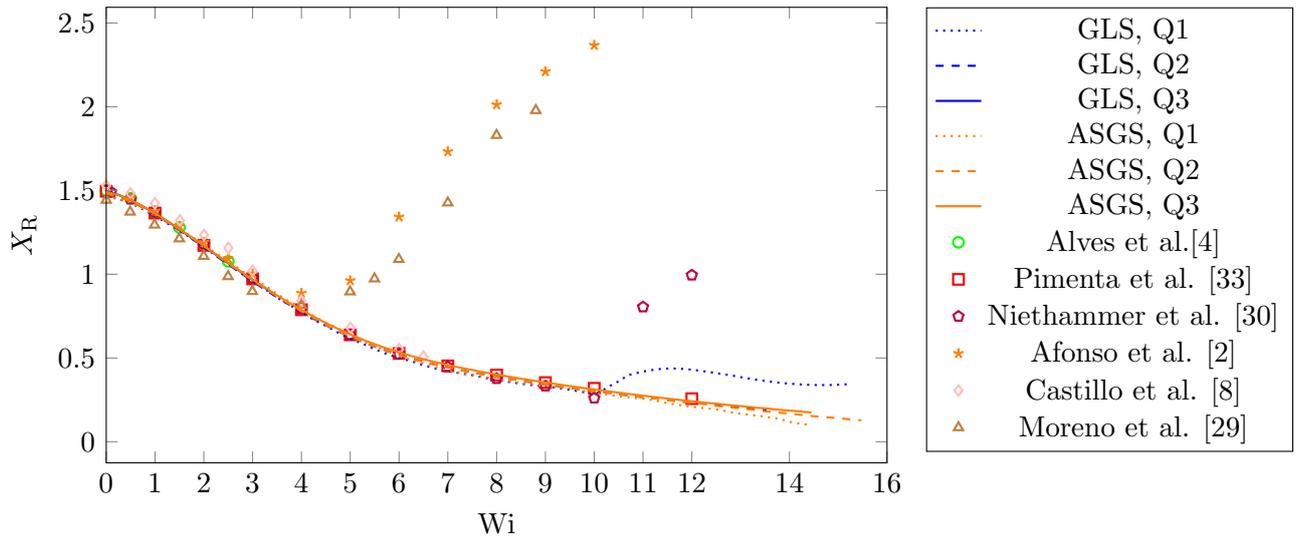
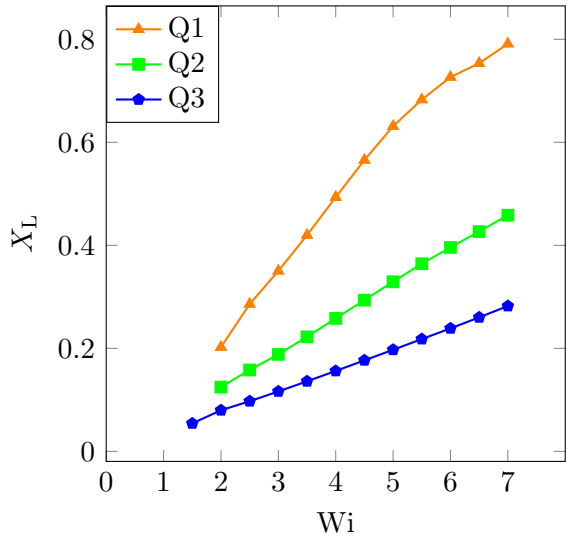
\begin{figure}[H]
	\begin{tikzpicture}
					\begin{axis}
					[
					set layers,					
					mark layer=axis background,
					width=0.45\textwidth,
				    height=0.45\textwidth,
					xlabel=\small{$\te{Wi}$},
					ylabel=\small{$X_{\te{L}}$},
					grid style=off,
					ticklabel style = {font=\small},
					legend style={at={(0.0,1.0)},anchor=north west,font=\small},
					xmin=0,
					xmax=8.0,
					xtick={0, 1.0, 2.0, 3.0, 4.0, 5.0, 6.0, 7.0},
					]

					\addplot[color = orange, mark=triangle*, thick, solid] table [x=Wi, y=VortexSize, col sep=comma] {fig/FourToOneContraction/VortexSize/LipVortexSize_ASS_Q1.csv};\addlegendentry{Q1};
					\addplot[color = green, mark=square*, thick, solid] table [x=Wi, y=VortexSize, col sep=comma] {fig/FourToOneContraction/VortexSize/LipVortexSize_ASS_Q2.csv};\addlegendentry{Q2};
					\addplot[color = blue, mark=pentagon*, thick, solid] table [x=Wi, y=VortexSize, col sep=comma] {fig/FourToOneContraction/VortexSize/LipVortexSize_ASS_Q3.csv};\addlegendentry{Q3};

					\end{axis}
					\end{tikzpicture}
\caption{Lip vortex sizes for the ASGS method of all mesh resolutions.}
\label{fig:4to1lipvortexlit}
\end{figure}	

\begin{table}[H]
\caption{Corner vortex size of ASGS method in comparison to literature values}
\label{tab:4to1cornervortexlit}
\begin{tabular}{llllllll}
$\Wi$ & Q1 & Q2 & Q3 & \makecell{Alves et al.\cite{Alves2003} \\ (Tab. 2)} & \makecell{Pimenta et al. \cite{Pimenta2017}\\(Tab. 2, \\In-house solver, M4)}& \cite{Niethammer2018} 4th root conf. \\
0.0  & 1.464 & 1.482 & 1.492 & 1.500 & 1.495  &                   \\
0.5  & 1.426 & 1.437 & 1.443 & 1.452 &            &  1.4483      \\
1.0  & 1.352 & 1.360 & 1.364 & 1.373 & 1.365  &   1.37154  \\
1.5  & 1.269 & 1.267 & 1.270 & 1.279 &             &                  \\
2.0  & 1.173 & 1.168 & 1.169 & 1.181 & 1.172   &  1.18034   \\
2.5  & 1.080 & 1.064 & 1.067 & 1.077 &             &                  \\
3.0  & 0.977 & 0.963 & 0.968 & 0.973 & 0.972   &  0.97722   \\
4.0  & 0.799 & 0.777 & 0.787 &          & 0.788    &  0.79316   \\
5.0  & 0.635 & 0.628 & 0.641 &          & 0.638    &  0.6382     \\
6.0  & 0.510 & 0.518 & 0.533 &          & 0.527    &  0.52204   \\
7.0  & 0.426 & 0.444 & 0.458 &          & 0.453    &  0.44667   \\
8.0  & 0.373 & 0.388 & 0.401 &          & 0.398    &  0.37626   \\
9.0  & 0.332 & 0.342 & 0.354 &          & 0.353    &  0.33182   \\
10.0 & 0.292 & 0.302 & 0.312 &          & 0.319    &  0.25996   \\
11.0 & 0.257 & 0.265 & 0.275 &          &              & 0.8041     \\
12.0 & 0.208 & 0.231 & 0.242 &          & 0.257   & 0.9943
\end{tabular}
\end{table}

\subsubsection{Stress profiles at high Weissenberg numbers}

\begin{figure}[H]		
					\subfigure[centerline $y=0$]{		
					\begin{tikzpicture}[scale=1.0]
					\begin{axis}
					[
					set layers,					
					mark layer=axis background,
					width=0.45\textwidth,
				    height=0.45\textwidth,
					xlabel=\small{$x$},
					ylabel=\small{$\sigma_{11}/ \br{\etaT \, \overline{u}_2 \, / \, H_2}$},
					grid style=off,
					ticklabel style = {font=\small},
					legend style={at={(1.0,1.0)},anchor=north east,font=\small},
					xmin=95,
					xmax=120,
					xtick={95, 100, 105, 110, 115, 120},
					xticklabels={-5, 0, 5, 10, 15, 20}
					]		
					\addplot[color = black, mark=none*, style=thin, solid] table [x=s, y=txx, col sep=comma] {fig/FourToOneContraction/CornerLineData/4to1Contraction_LineplotCenter_ASS_Q3_Wi1.0000.csv};\addlegendentry{$\Wi=1.0$};		
					
					\addplot[color = green, mark=none*, style=thin, solid] table [x=s, y=txx, col sep=comma] {fig/FourToOneContraction/CornerLineData/4to1Contraction_LineplotCenter_ASS_Q3_Wi2.0000.csv};\addlegendentry{$\Wi=2.0$};		
					
					\addplot[color = magenta, mark=none*, style=thin, solid] table [x=s, y=txx, col sep=comma] {fig/FourToOneContraction/CornerLineData/4to1Contraction_LineplotCenter_ASS_Q3_Wi3.0000.csv};\addlegendentry{$\Wi=3.0$};	

					\addplot[color = blue, mark=none*, style=thin, solid] table [x=s, y=txx, col sep=comma] {fig/FourToOneContraction/CornerLineData/4to1Contraction_LineplotCenter_ASS_Q3_Wi5.0000.csv};\addlegendentry{$\Wi=5.0$};		
					
					\addplot[color = red, mark=none*, style=thin, solid] table [x=s, y=txx, col sep=comma] {fig/FourToOneContraction/CornerLineData/4to1Contraction_LineplotCenter_ASS_Q3_Wi7.0000.csv};\addlegendentry{$\Wi=7.0$};		
					
					\addplot[color = orange, mark=none*, style=thin, solid] table [x=s, y=txx, col sep=comma] {fig/FourToOneContraction/CornerLineData/4to1Contraction_LineplotCenter_ASS_Q3_Wi9.0000.csv};\addlegendentry{$\Wi=9.0$};

					\addplot[color = black, mark=none*, style=thin, dashed] table [x=s, y=txx, col sep=comma] {fig/FourToOneContraction/CornerLineData/4to1Contraction_LineplotCenter_ASS_Q2_Wi1.0000.csv};
					
					\addplot[color = green, mark=none*, style=thin, dashed] table [x=s, y=txx, col sep=comma] {fig/FourToOneContraction/CornerLineData/4to1Contraction_LineplotCenter_ASS_Q2_Wi2.0000.csv};
					
					\addplot[color = magenta, mark=none*, style=thin, dashed] table [x=s, y=txx, col sep=comma] {fig/FourToOneContraction/CornerLineData/4to1Contraction_LineplotCenter_ASS_Q2_Wi3.0000.csv};

					\addplot[color = blue, mark=none*, style=thin, dashed] table [x=s, y=txx, col sep=comma] {fig/FourToOneContraction/CornerLineData/4to1Contraction_LineplotCenter_ASS_Q2_Wi5.0000.csv};
					
					\addplot[color = red, mark=none*, style=thin, dashed] table [x=s, y=txx, col sep=comma] {fig/FourToOneContraction/CornerLineData/4to1Contraction_LineplotCenter_ASS_Q2_Wi7.0000.csv};
					
					\addplot[color = orange, mark=none*, style=thin, dashed] table [x=s, y=txx, col sep=comma] {fig/FourToOneContraction/CornerLineData/4to1Contraction_LineplotCenter_ASS_Q2_Wi9.0000.csv};
					\end{axis}
					\end{tikzpicture}	
					}~
					\subfigure[cornerline $y=-H_2$]{		
					\begin{tikzpicture}[scale=1.0]
					\begin{axis}
					[
					set layers,					
					mark layer=axis background,
					width=0.45\textwidth,
				    height=0.45\textwidth,
					xlabel=\small{$x$},
					ylabel=\small{$\sigma_{11}/ \br{\etaT \, \overline{u}_2 \, / \, H_2}$},
					grid style=off,
					ticklabel style = {font=\small},
					legend style={at={(1.0,1.0)},anchor=north east,font=\small},
					xmin=99,
					xmax=103,
					xtick={99, 100, 101, 102, 103},
					xticklabels={-1, 0, 1, 2, 3}
					]		
					\addplot[color = black, mark=none*, style=thin, solid] table [x=s, y=txx, col sep=comma] {fig/FourToOneContraction/CornerLineData/4to1Contraction_LineplotCorner_ASS_Q3_Wi1.0000.csv};
					
					\addplot[color = green, mark=none*, style=thin, solid] table [x=s, y=txx, col sep=comma] {fig/FourToOneContraction/CornerLineData/4to1Contraction_LineplotCorner_ASS_Q3_Wi2.0000.csv};
					
					\addplot[color = magenta, mark=none*, style=thin, solid] table [x=s, y=txx, col sep=comma] {fig/FourToOneContraction/CornerLineData/4to1Contraction_LineplotCorner_ASS_Q3_Wi3.0000.csv};

					\addplot[color = blue, mark=none*, style=thin, solid] table [x=s, y=txx, col sep=comma] {fig/FourToOneContraction/CornerLineData/4to1Contraction_LineplotCorner_ASS_Q3_Wi5.0000.csv};
					
					\addplot[color = red, mark=none*, style=thin, solid] table [x=s, y=txx, col sep=comma] {fig/FourToOneContraction/CornerLineData/4to1Contraction_LineplotCorner_ASS_Q3_Wi7.0000.csv};
					
					\addplot[color = orange, mark=none*, style=thin, solid] table [x=s, y=txx, col sep=comma] {fig/FourToOneContraction/CornerLineData/4to1Contraction_LineplotCorner_ASS_Q3_Wi9.0000.csv};

					\addplot[color = black, mark=none*, style=thin, dashed] table [x=s, y=txx, col sep=comma] {fig/FourToOneContraction/CornerLineData/4to1Contraction_LineplotCorner_ASS_Q2_Wi1.0000.csv};
					
					\addplot[color = green, mark=none*, style=thin, dashed] table [x=s, y=txx, col sep=comma] {fig/FourToOneContraction/CornerLineData/4to1Contraction_LineplotCorner_ASS_Q2_Wi2.0000.csv};
					
					\addplot[color = magenta, mark=none*, style=thin, dashed] table [x=s, y=txx, col sep=comma] {fig/FourToOneContraction/CornerLineData/4to1Contraction_LineplotCorner_ASS_Q2_Wi3.0000.csv};

					\addplot[color = blue, mark=none*, style=thin, dashed] table [x=s, y=txx, col sep=comma] {fig/FourToOneContraction/CornerLineData/4to1Contraction_LineplotCorner_ASS_Q2_Wi5.0000.csv};
					
					\addplot[color = red, mark=none*, style=thin, dashed] table [x=s, y=txx, col sep=comma] {fig/FourToOneContraction/CornerLineData/4to1Contraction_LineplotCorner_ASS_Q2_Wi7.0000.csv};
					
					\addplot[color = orange, mark=none*, style=thin, dashed] table [x=s, y=txx, col sep=comma] {fig/FourToOneContraction/CornerLineData/4to1Contraction_LineplotCorner_ASS_Q2_Wi9.0000.csv};
					\end{axis}
					\end{tikzpicture}	
					}
\caption{Stress profiles in the 4:1 contraction for mesh Q3 (solid) and Q2 (dashed) using the ASGS method of various Weissenberg numbers.}
\label{fig:4to1line}
\end{figure}
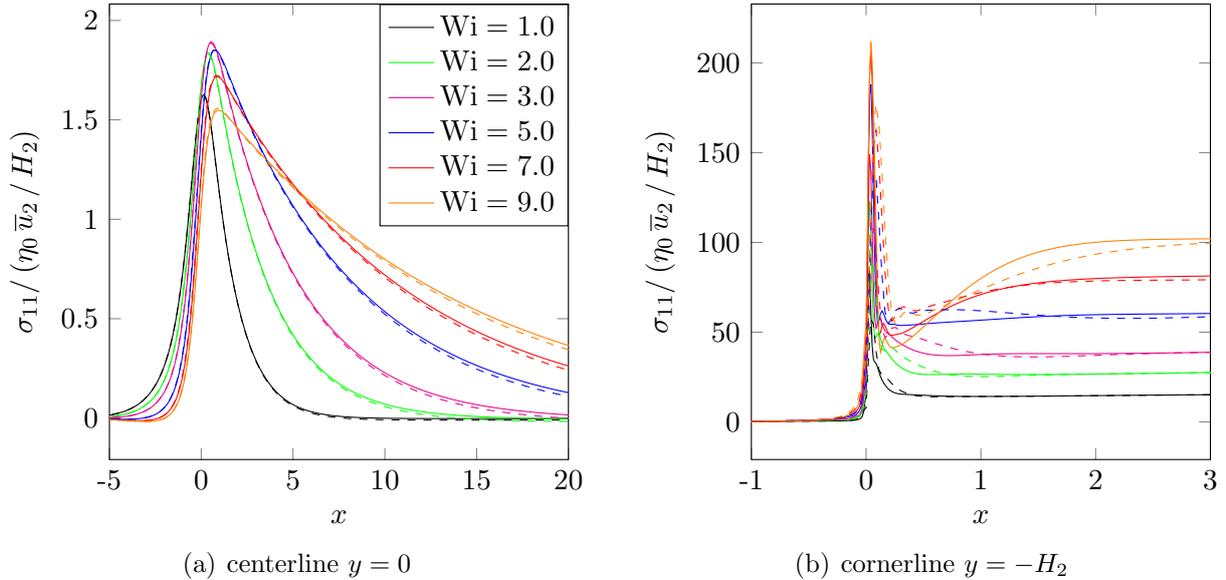

Lastly, we present values for the first normal stress component $\sigma_{11}$ along the centerline $y=0$ in Fig.~\ref{fig:4to1line}~(a), and along the re-entrant corner $y=-H_2$ in Fig.~\ref{fig:4to1line}~(b). The peak values at the re-entrant corner increase with the elasticity of the fluid. Differences in the peak values and downstream are visible, indicating the stress boundary layer is not completely resolved. As pointed out in many previous works, e.g., \cite{Niethammer2018}, the stresses can not be expected to converge at the re-entrant corner due to the geometric singularity. Finer meshes will increase the peak values. We could not find matching data in the literature.
The profiles in the centerline agree well with the presented meshes. The solution is smooth in this region. The maximum values of stresses and velocity match well with table-based data by Alves \cite{Alves2003}, see Tab.~\ref{tab:4to1maxstress} and Tab.~\ref{tab:4to1maxvelo}.

\begin{minipage}[c]{0.45\textwidth}
\centering
\begin{table}[H]
\centering
\caption{Maximum values of first normal stress component $\chi_{11}^{\te{max}}$ along the centerline ${y=0}$.}
\begin{tabular}{c|c|c|c}
$\Wi$ & GLS & ASGS & \makecell{Alves et al. \cite{Alves2003} \\ (Table 2)}\\ 
\hline 
0.0 & 0.320 & 0.320 & 0.360 \\ 
0.5 & 0.460 & 0.460 & 0.461 \\
1.0 & 0.541 & 0.542 & 0.544 \\
1.5 & 0.587 & 0.587 & 0.589 \\ 
2.0 & 0.611 & 0.612 & 0.612 \\
2.5 & 0.624 & 0.625 & 0.623 \\
3.0 & 0.629 & 0.631 & 0.638
\end{tabular}
\label{tab:4to1maxstress}
\end{table}
\end{minipage}
\hspace{0.05\textwidth}
\begin{minipage}[c]{0.45\textwidth}
\centering
\begin{table}[H]
\centering
\caption{Maximum values of first velocity component $u_1^{\te{max}}$ along the centerline ${y=0}$.}
\begin{tabular}{c|c|c|c}
$\Wi$ & GLS & ASGS & \makecell{Alves et al. \cite{Alves2003} \\ (Table 2)}\\ 
\hline 
0.0 & 1.501 & 1.501 & 1.501 \\
0.5 & 1.511 & 1.511 & 1.511 \\
1.0 & 1.525 & 1.526 & 1.525 \\
1.5 & 1.537 & 1.537 & 1.537 \\
2.0 & 1.547 & 1.547 & 1.546 \\
2.5 & 1.556 & 1.556 & 1.554 \\
3.0 & 1.563 & 1.563 & 1.562
\end{tabular}
\label{tab:4to1maxvelo}
\end{table}
\end{minipage}

\section{Conclusions}
\label{sec:conclusion}

In this work, stabilized methods for an alternate fully-implicit log-conf representation for the Oldroyd-B model were proposed. The non-singular nature of this law allows the use of continuation techniques starting in the Newtonian limit. The methods include SUPG and GLS methods and the ASGS method based on the VMS framework.

We found a second-order spatial accuracy of all methods in a four-roll periodic box. The methods were further validated with a flow past a cylinder and a flow through a 4:1 contraction. We could see good agreement in the results for all methods with literature data below a critical Weissenberg number, where discrepancies become larger. In the 4:1 contraction benchmark, the results revealed a pronounced mesh dependence near the re-entrant corner for highly elastic flows. In particular, the lip vortex was shown to be a highly mesh-sensitive quantity. The corner vortex, on the other hand, displayed only a slight dependence. Our corner vortex predictions match well with previously published results up to $\Wi=9$. Finer resolutions would be required to obtain a definite answer to the prediction of lip vortex sizes. However, the element size near the re-entrant corner posed strict limitations on the limiting Weissenberg number. The use of divergence-free velocity interpolation spaces may improve the robustness. A non-zero divergence can introduce a source term in the constitutive law, which might cause an unbounded growth of the stresses. The proposed methods are also only weakly consistent due to the deletion of seconder order derivatives. A globally consistent scheme may improve accuracy.

While we observed only minor differences between the methods in predicting flow parameters, there were significant differences in the highest possible computable Weissenberg number. The ASGS method was shown to produce results for higher Weissenberg numbers than the GLS or SUPG approach. Analytic Jacobians, where the stabilization parameters were treated in a fixed-point manner, showed sub-optimal convergence and overall reduced robustness for more elastic flows. The use of finite differences for approximating the Jacobians --- including non-linearities from the stabilization parameters --- seemed to be an essential constituent concerning the robustness of the methods.

Although the steady-state problem was solved, the unique combination of a fully implicit logarithmic conformation formulation, algebraic sub-grid scales, and central finite differences resulted in a robust methodology, confirming and contributing to the results with the highest Weissenberg numbers reported in the literature.


\section{Acknowledgments}

This work was funded by the Deutsche Forschungsgemeinschaft (DFG, German Research 
Foundation) – 333849990/GRK2379 (IRTG Modern Inverse Problems).
Computing resources were provided by the Jülich-Aachen Research Alliance JARA-HPC. 
The authors gratefully acknowledge the computing time granted by the JARA Vergabegremium and provided on the JARA Partition part of the supercomputer CLAIX at RWTH Aachen University.

\bibliographystyle{abbrvnat}
\bibliography{biblib/literature} 
\end{document}